\documentclass{aa}  

\usepackage[caption = false]{subfig}
\usepackage{float}
\usepackage{graphicx}
\usepackage{booktabs,threeparttable}
\usepackage{threeparttablex} 
\usepackage{amsmath}
\usepackage{longtable}
\usepackage{subfig}
\usepackage{xcolor}
\usepackage{placeins} 
\usepackage{txfonts}

\usepackage{hyperref}
\hypersetup{
    colorlinks=true,
    linkcolor=blue,filecolor=magenta, citecolor=blue,urlcolor=blue
}

\makeatletter
\renewcommand*\aa@pageof{, page \thepage{} of \pageref*{LastPage}}
\makeatother

\usepackage[capitalise]{cleveref} 

\newcommand{\kms}{\ensuremath{\rm km\,s^{-1}}}
\newcommand{\Msol}{\ensuremath{M_{\odot}}}
\newcommand{\Lsol}{\ensuremath{L_{\odot}}}

\begin{document}

   \title{Core to ultracompact \textsc{Hii} region evolution in the W49A massive protocluster}
    
   \author{T. Nony\inst{1,2}
          \and R. Galván-Madrid\inst{2} 
          \and N. Brouillet\inst{3}  
          \and G. Suárez\inst{4}    
          \and F. Louvet\inst{5}     
          \and C. G. De Pree\inst{6}  
          \and M. Juárez-Gama\inst{7} 
          \and A. Ginsburg\inst{8}   
          \and K. Immer\inst{9}   
          \and Y. Lin\inst{10}  
          \and H. B. Liu\inst{11,12}  
          \and C. G. Román-Zúñiga\inst{13}   
          \and Q. Zhang\inst{14}      
          }

   \institute{INAF - Osservatorio Astrofisico di Arcetri, Largo E. Fermi 5, 50125 Firenze, Italy
    \email{thomas.nony@inaf.it}
    \and Instituto de Radioastronomía y Astrofísica, Universidad Nacional Autónoma de México, Morelia, Michoacán 58089, Mexico 
    \and Laboratoire d’astrophysique de Bordeaux, Univ. Bordeaux, CNRS, B18N, allée Geoffroy Saint-Hilaire, 33615 Pessac, France
    \and Department of Astrophysics, American Museum of Natural History, Central Park West at 79th Street, NY 10024, USA
    \and Univ. Grenoble Alpes, CNRS, IPAG, 38000 Grenoble, France
    \and National Radio Astronomy Observatory, 520 Edgemont Road, Charlottesville, VA 22903
    \and Instituto Nacional de Astrofísica, Óptica y Electrónica, Luis E. Erro 1, 72840 Tonantzintla, Puebla, Mexico
    \and Department of Astronomy, University of Florida, PO Box 112055, USA 
    \and Leiden Observatory, Leiden University, PO Box 9513, 2300 RA Leiden, The Netherlands 
    \and Max-Planck-Institut für Extraterrestrische Physik, Giessenbachstr. 1, 85748 Garching bei München, Germany 
    \and Department of Physics, National Sun Yat-Sen University, No. 70, Lien-Hai Road, Kaohsiung City 80424, Taiwan, R.O.C. 
    \and Center of Astronomy and Gravitation, National Taiwan Normal University, Taipei 116, Taiwan 
    \and Universidad Nacional Autónoma de México, Instituto de Astronomía, AP 106, Ensenada 22800, BC, Mexico 
    \and Center for Astrophysics, Harvard \& Smithsonian, 60 Garden St., Cambridge, MA 02420, USA} 

 
  \abstract
   {}
   {We aim to identify and characterize cores in the high-mass proto-cluster W49, determine their evolutionary stages and measure the associated lifetimes.}
   {We built a catalog of 129 cores extracted from an ALMA 1.3~mm continuum image at 0.26$\arcsec$ (2900~au) angular resolution. The association between cores and Hyper/Ultra Compact \textsc{Hii} (H/UC\,\textsc{Hii}) regions was established from the analysis of VLA 3.3~cm continuum and  H30$\alpha$ line observations. We also looked for emission of hot molecular cores (HMCs) using the methyl formate doublet at 218.29~GHz.}
   {We identified 40 cores associated with an H/UC\,\textsc{Hii} region and 19 HMCs over the ALMA mosaic. The 52 cores with an H/UC\,\textsc{Hii} region and/or a HMC are assumed to be high-mass protostellar cores, 
   while the rest of the core population likely consists in prestellar cores and low-mass protostellar cores. We found a good agreement between the two tracers of ionized gas, with 23 common detections and only four cores detected at 3.3~cm and not in H30$\alpha$. The spectral indexes from 3.3~cm to 1.3~mm range from 1, for the youngest cores with partially optically thick free-free emission, to about -0.1, that is optically thin free-free emission obtained for cores likely more evolved.}
   {Using as a reference the H/UC\,\textsc{Hii} regions, we found the statistical lifetimes of the HMC and massive protostellar phases in W49N to be about $6\times10^4$\,yr and $1.4\times10^5$\,yr, respectively. We also showed that HMC can co-exist with H/UC\,\textsc{Hii} regions during a short fraction of the core lifetime, about 2$\times10^4$~yr. This indicates a rapid dispersal of the inner molecule envelope once the HC\,\textsc{Hii} is formed.} 


   \maketitle
%

\section{Introduction}
The evolutionary sequence  and timescales of the formation of high-mass stars are still poorly constrained \citep[see e.g., the review by][]{2018Motte}, partly because of observational difficulties related to low-number statistics and large distances within the Milky Way. W49A is among the few molecular clouds that enable us to study, in a single region and with robust statistics, various evolutionary stages of high-mass star formation. 
Young, massive protostars first drive strong molecular outflows and heat the inner part of their envelope. Complex organic molecules (COMs) are released from the grains in the heated region \citep{2009ARA&A..47..427H}, which is called hot molecular core (HMC) and is thus characterized by multiple lines from COMs - the so called "line forest". Contrarily to their low-mass analogs, high-mass protostars are still embedded in their nascent envelope when they evolve to the main sequence. The extreme ultraviolet (EUV) radiation from high-mass stars (>8\,$\Msol$) is sufficient to ionize hydrogen and create an \textsc{Hii} region \citep{2002ARA&A..40...27C}. As the luminosity of the star keeps increasing, the \textsc{Hii} region expands and breaks through the core, and eventually impacts the whole molecular cloud. 
Hypercompact (HC) \textsc{Hii} regions are usually interpreted as the youngest stage of ionization by a massive star \citep[e.g.,][]{2007ApJ...666..976K, 2016ApJ...818...52T}, confined to within cores to sizes $\sim 10^{-2}$ pc, or even less, and having electron densities $n_e > 10^6$ cm$^{-3}$ \citep{Kurtz05}. Ultracompact (UC) \textsc{Hii} regions are, by definition, expected to be larger and less dense by at least one order of magnitude. However, deep centimeter surveys of protocluster regions \citep[e.g.,][]{2016A&A...595A..27G} have shown that the most common type of hypercompact sized \textsc{Hii} regions have lower densities, in the range $10^4$ to $<10^6$ cm$^{-3}$ \citep{2020ApJ...899...94R}. Since a reliable distinction between these two types of objects requires modeling of the cm to (sub)mm spectral energy distribution \citep[e.g.,][]{2008ApJ...672..423K,2009ApJ...706.1036G,2022ApJ...936...68Z}, we refer to all our detections of small \textsc{Hii} emission as H/UC\,\textsc{Hii} regions.

W49A, located at a distance of 11.1~kpc from the Sun \citep{2013ApJ...775...79Z}, is one of the most luminous protoclusters in the Galaxy ($L \sim 2.6\times10^{7}\Lsol$, \citealt{1991A&A...251..231S,2016ApJ...828...32L}). 
The W49A giant molecular cloud (GMC) has been mapped from $\sim 100$ pc to sub-pc scales in the (sub)millimeter continuum and in molecular lines, down to angular resolutions of a few arcseconds \citep[e.g.,][]{2009PASJ...61...39M, 2010A&A...520A..84P,2013ApJ...779..121G,2020MNRAS.497.1972}. The W49A GMC can be divided in three subcomponents labeled W49 north (W49N), W49 south (W49S), and W49 south-west (W49SW).
From CO observations, \citet{2013ApJ...779..121G} evaluated the gas mass of the W49A GMC within a radius of 60 pc to be $10^6\,\Msol$, with the main star-forming hub W49N concentrating $M_{\rm gas} \sim 2\times10^5\,\Msol$ within a radius of 6 pc. 
A cluster of massive stars with a total mass of about $10^4\,\Msol$ has been identified in the near infrared by \cite{2005A&A...430..481H}, at about 3~pc east of the center of W49N. The study of \cite{2015ApJ...813...25S} with \textit{Spitzer} also revealed 232 Class 0/I Young Stellar Objects in the entire W49A GMC.
About 50 HC and UC\,\textsc{Hii} regions have been detected in W49A \citep[e.g.,][]{1987Sci...238.1550W,1997ApJ...482..307D,2020AJ....160..234D}. The embedded and infrared-visible stellar populations drive already significant feedback across the GMC \citep{2019A&A...622A..48R}. 
Prominent water maser features extending to hundreds of kilometers per second from the cloud velocity have been reported within W49N \citep[e.g.,][]{1992ApJ...393..149G,2004ApJS..155..577M}.
Also, five 6.7\,GHz methanol masers were reported the catalog of \cite{2015MNRAS.450.4109B}. 
Finally, \cite{2001ApJ...550L..81W} reported the detection of 6 HMCs in W49N using CH$_{3}$CN observations at subarcsecond resolution, and \cite{2022Miyawaki} studied the HMC W49N MCN-a (UC\textsc{Hii} region J1 in \citealt{1997ApJ...482..307D}). 

The star forming core population and star formation sequence in W49A have however not been explored in detail so far. In this work, we present a subarcsecond resolution study of W49A aimed at identifying and characterizing its core populations, as well as the evolutionary sequence from young star forming cores to UC\textsc{Hii} regions.  
The ALMA and VLA observations we used are presented in \cref{s:obs}. The  analysis of the continuum maps and molecular lines presented in \cref{s:analy} is used in \cref{s:charac} to characterize the dust cores and establish a temporal classification based on their association with hot molecular cores and UC~\textsc{Hii} regions. A discussion is proposed in \cref{s:discu} and \cref{s:conclu} summarizes our main conclusion.

\section{Observations}
\label{s:obs}

\subsection{ALMA data}
\label{su:ALMAdat}

Observations of W49A at 1.3~mm (Band 6) were carried out with ALMA in Cycle 5 (project 2016.1.00620.S, PIs: Ginsburg, Galván-Madrid) between 2017 and 2018.
The data consist in a mosaic of 28 fields with a primary beam of 25.6$\arcsec$, for a total mapping extent of about $2.5\arcmin \times 1.6\arcmin$ ($8.2 \times 5.3$~pc). The maximum recoverable scale of the 12-m array configuration used in this work is about 11$\arcsec$.

Continuum data were processed with \texttt{CASA} version 5.4 \citep{2022CASA} using the imaging and self-calibration pipeline developed by the ALMA-IMF consortium (described in detail in \citealt{Ginsburg22}). The continuum map has been produced using a selection of line-free channels (equivalent to the \texttt{cleanest} maps in \citealt{Ginsburg22}) from the four spectral windows (hereafter Spws) of the dataset, summing up to a bandwidth of 1.23~GHz.
We performed three iterations of cleaning and phase self-calibration and one iteration of amplitude self-calibration, using model masks of increasing size and decreasing cleaning thresholds We used the
Multi-term (Multi Scale) Multi-Frequency Synthesis deconvolver of \texttt{tclean}
with two Taylor terms, a robust parameter of 0, and scales up to nine times larger than the synthesized beam.
The final beam of the 1.3~mm continuum image has a FWHM size of $0.29\arcsec \times 0.24\arcsec$ ($\approx 2900$\,au at 11.1\,kpc). The rms noise measured in regions far away from the brightest emission, where the map is not limited by dynamic range, is of 0.38 mJy\,beam$^{-1}$ (see \cref{tab:data}).

For the line data,  the four spectral windows were processed with \texttt{IMAGER}\footnote{https://imager.oasu.u-bordeaux.fr}, implemented within the \texttt{GILDAS} software\footnote{http://www.iram.fr/IRAMFR/GILDAS}. Cleaning was performed using Clark deconvolution and 0.1 robust  weighting. The cubes across the four bands have a spatial resolution of approximately 0.24$\arcsec$ (2750\,au), a channel width of 0.65 km\,s$^{-1}$, and an rms noise of 1.6 mJy\,beam$^{-1}$ (0.6\,K) per channel. The detailed parameters for each spectral window are given in \cref{tab:data}.  
We have used the method presented in \citet{2019Molet} and \citet{2022Brouillet} to separate the continuum and spectral line emission in each pixel of the image plane. 
The Spw~2 band covers the H30$\alpha$ line at 231.9~GHz. We also used the Spw~0 and Spw~3 bands to analyze CH$_3$OCHO lines.

\subsection{VLA data}
\label{su:VLAdat}

\begin{figure*}
    \centering
    \includegraphics[width=\hsize]{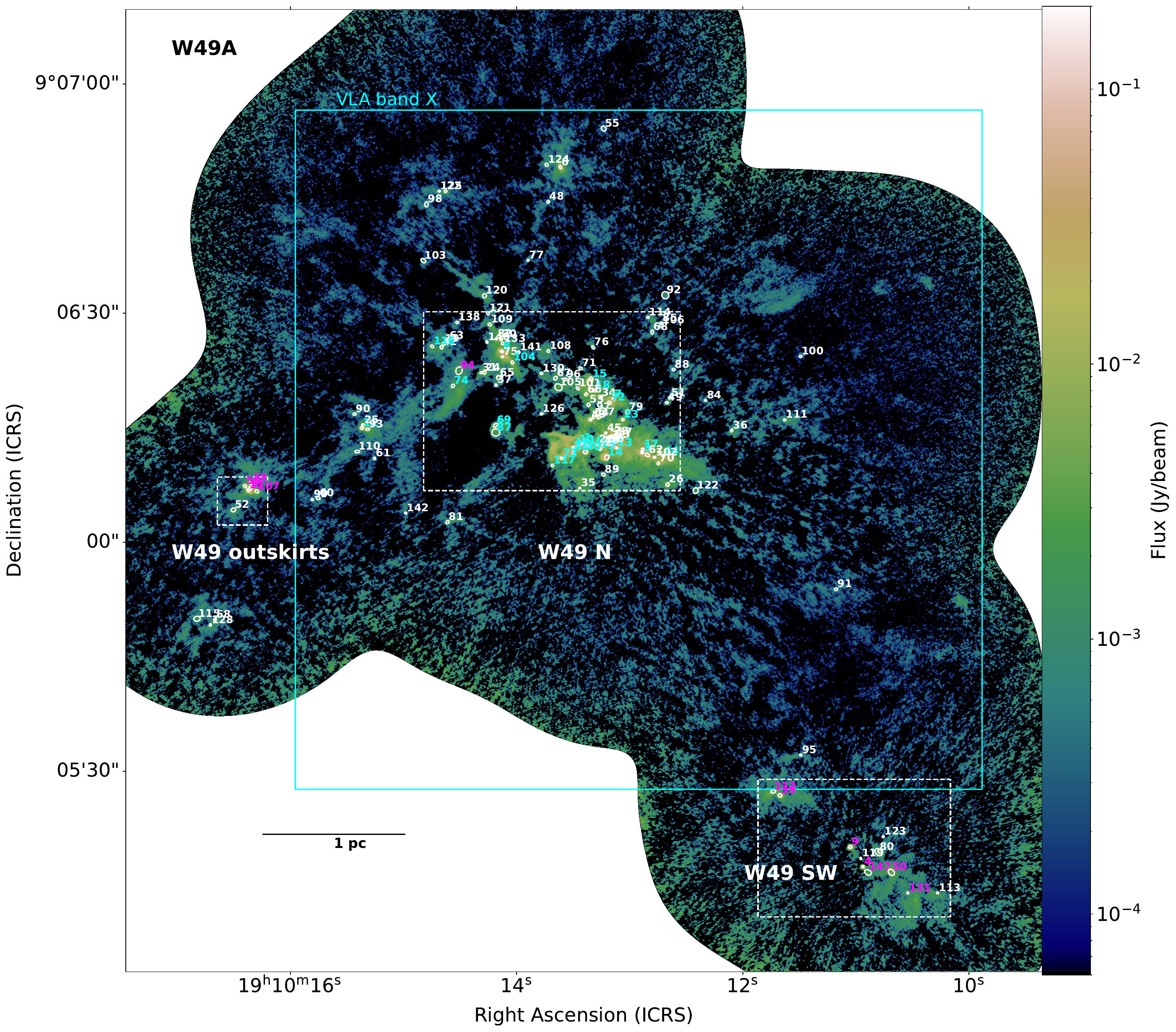}
    \caption{The W49A protocluster region as imaged at 1.3~mm using ALMA. 
    Dust cores are represented by their FWHM ellipses in white, and numbered in cyan for cores also detected at 3.3\,cm, in magenta for cores detected only in H30$\alpha$, and white for non detections at 3.3\,cm nor H30$\alpha$. The field of view of the VLA X band image is represented in solid cyan line. A  scale bar is shown. Zoom-ins toward the three main regions of interest (W49N, W49SW and W49 outskirts), delimited by white dashed lines, are shown in \cref{fig:zomm-center,fig:zomm-swout}.}
    \label{fig:contB6}
\end{figure*}

We used the VLA A-array continuum image at 3.3~cm (X Band) presented in \cite{2020AJ....160..234D}. This image was created with the purpose of only studying W49N. Its field of view is outlined in Fig.~\ref{fig:contB6}.
We smoothed this image from its original angular resolution (0.16$\arcsec$) to the resolution of the B6 ALMA continuum image (0.26$\arcsec$), then corrected it for its primary-beam response and regridded it to the ALMA image frame.

Juárez-Gama et al. (in prep.) used this image to provide a dendrogram catalog of H/UC\,\textsc{Hii} regions and their physical properties in W49N. Their catalog of 79 H/UC\,\textsc{Hii}s is consistent with, and more complete, than previous visual identifications of H/UC\,\textsc{Hii}s in this protocluster \citep[e.g.,][]{2005ApJ...624L.101D,2020AJ....160..234D}. 
We used the cm continuum catalog of Juárez-Gama et al. to complement the identification of dust cores and HMCs reported in this paper, as well as the \textsc{Hii} emission traced by H$30\alpha$ also presented in this paper. 
 
\begin{table*}[htb]
   \begin{threeparttable}
    \caption[]{\label{tab:data}Parameters of the continuum images and spectral cubes.}
    \begin{tabular}{ c c c c c c c c c }
    \hline
    \hline
     Image   &  Instr.  &   $\nu_{\rm obs}$\tnote{a}  & Bandwidth\tnote{b}  & pixel & \multicolumn{2}{c}{Resolution\tnote{c}}  & $1\sigma$ rms \\
       &    &  [GHz]        &   [MHz]    & [$\arcsec$] & [$\arcsec \times \arcsec$] & [km~s$^{-1}$]  & [mJy\,beam$^{-1}$]\tnote{d} \\
     \hline
    X band cont. & VLA & 9.107  &  688  & 0.06 & $0.29 \times 0.24$  & -  & 0.22 \\
    Band 6 cont. &  ALMA  & 227.950  &  1230  & 0.06 & $0.29 \times 0.24$  & -  & 0.38 \\
    Spw 0 & ALMA & 217.900   & 1875  & 0.07 & $0.28 \times 0.23$  & 0.67 & 1.47 \\ 
    Spw 1 & ALMA  & 219.860  & 1875  & 0.07 & $0.28 \times 0.25$ & 0.67 & 1.67 \\
    Spw 2/H30$\alpha$ & ALMA &  231.870   & 1875  & 0.07 & $0.27 \times 0.20$  & 0.63 & 1.57 \\ 
    Spw 3 & ALMA  & 233.740  & 1875  & 0.07 & $0.27 \times 0.20$ & 0.63  & 1.57 \\
    \hline
    \end{tabular} 
        \begin{tablenotes}
\item [a] Central frequency of the continuum images or of the spectral windows. The central frequency of the ALMA continuum image is computed as the intensity-weighted average frequency, with a spectral index of 3 (see Eq. 1 in \citealt{Ginsburg22}). 
\item [b] Bandwidth used to create the continuum images or bandwidth of the cubes.
\item [c] Beam FWHM and channel width. The data cubes have been produced using the default Hanning smoothing function without channel averaging (N=1), resulting in a spectral resolution of two channels. 
\item [d] 1 mJy\,beam$^{-1}$ corresponds to 0.34~K at 230 GHz and to 222~K at 9 GHz for the continuum maps.
    \end{tablenotes}
\end{threeparttable}
\end{table*}

\section{Analysis of the continuum maps and molecular lines}
\label{s:analy}

\begin{figure*}
    \centering
    \includegraphics[width=0.7\hsize]{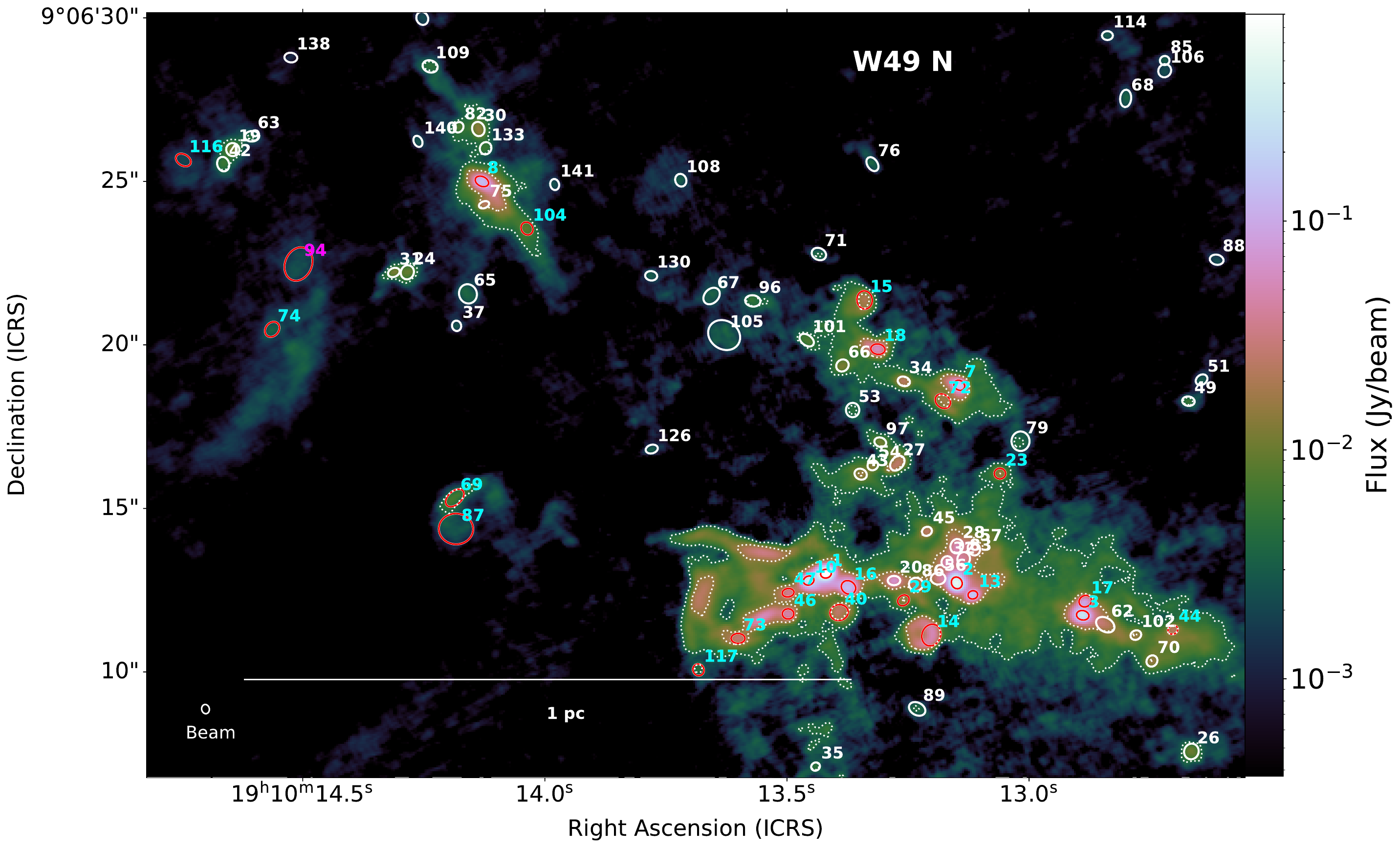}
    \includegraphics[width=0.7\hsize]{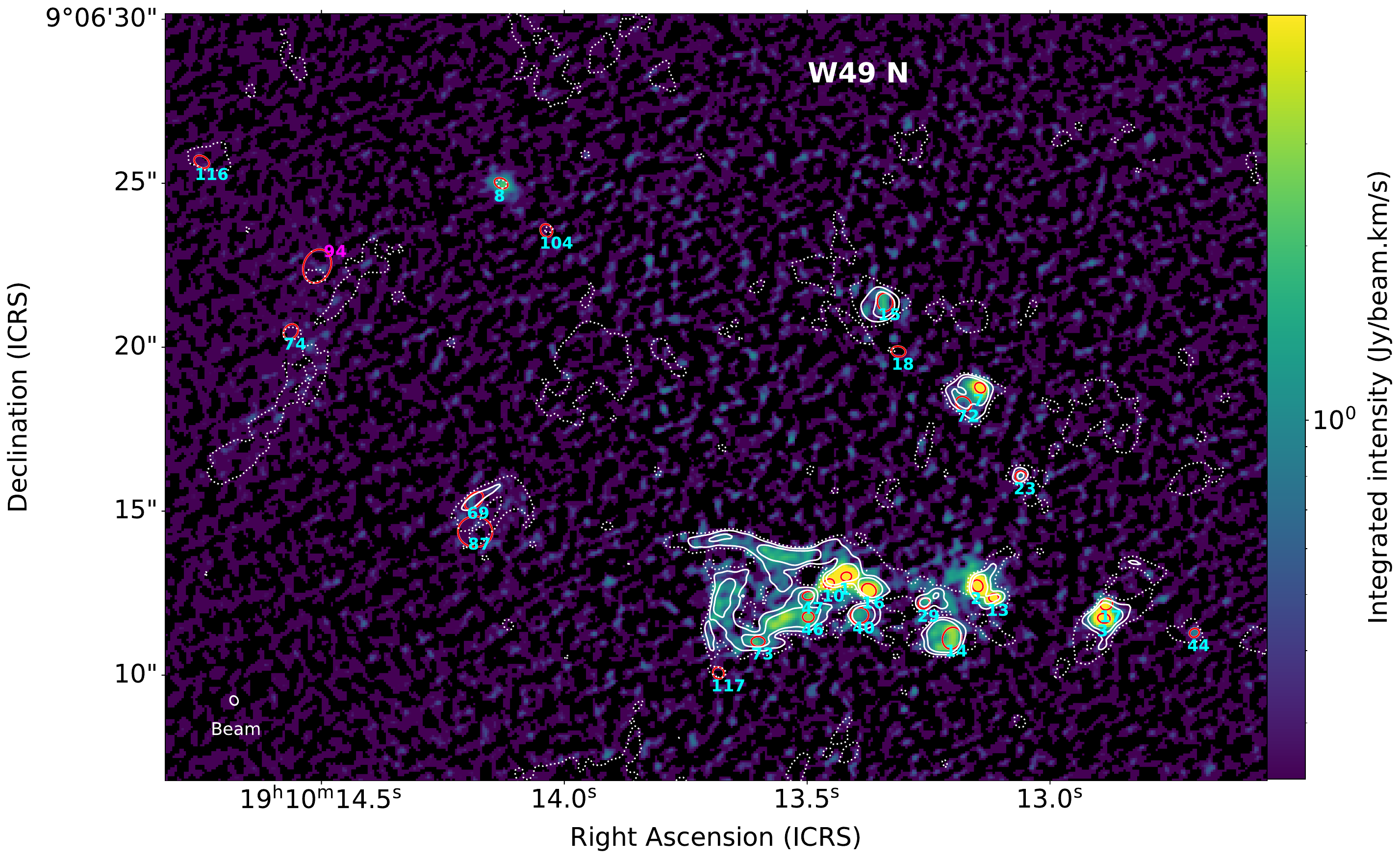}
    \caption{1.3 millimeter continuum and recombination line maps toward W49N at the center of W49A. Top panel: ALMA B6 continuum image, contours are 10 and 40 sigma. Bottom panel: moment 0 map of H30$\alpha$ (color) and VLA X-band map (contours). Contours are 3 sigma (dotted lines), 15, and 40 sigma (solid lines). Dust cores are represented by their FWHM ellipses and numbered with the same colors as in \cref{fig:contB6}. Cores detected in X band and/or in H30$\alpha$ are shown in red, the top panel also shows the remainder of the core catalog in white. Beams are shown in the lower left corner. A scale bar is shown in the top panel.}
    \label{fig:zomm-center}
\end{figure*}

\begin{figure*}
    \subfloat{\includegraphics[width=0.55\hsize]{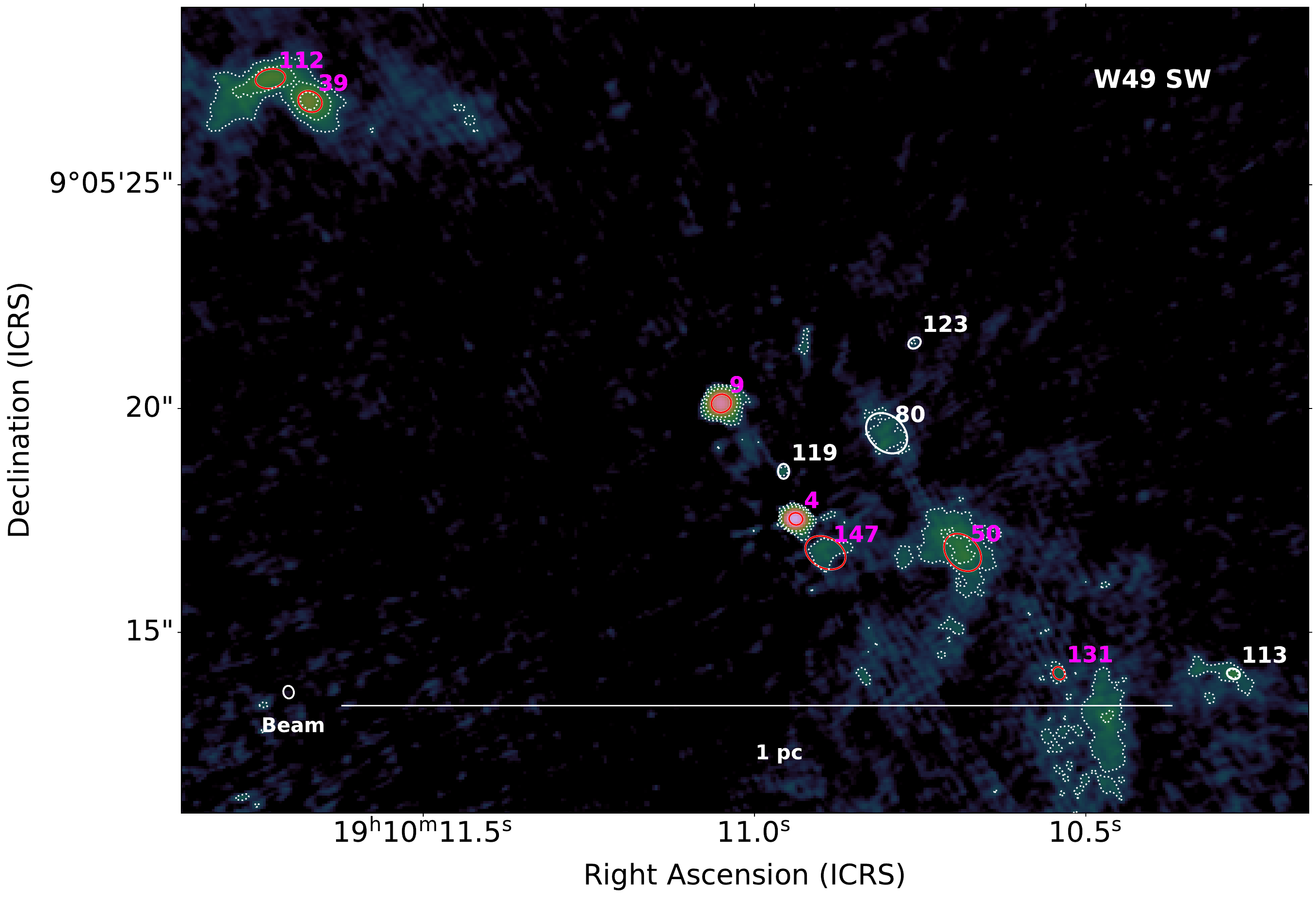}}
    \subfloat{\includegraphics[width=0.42\hsize]{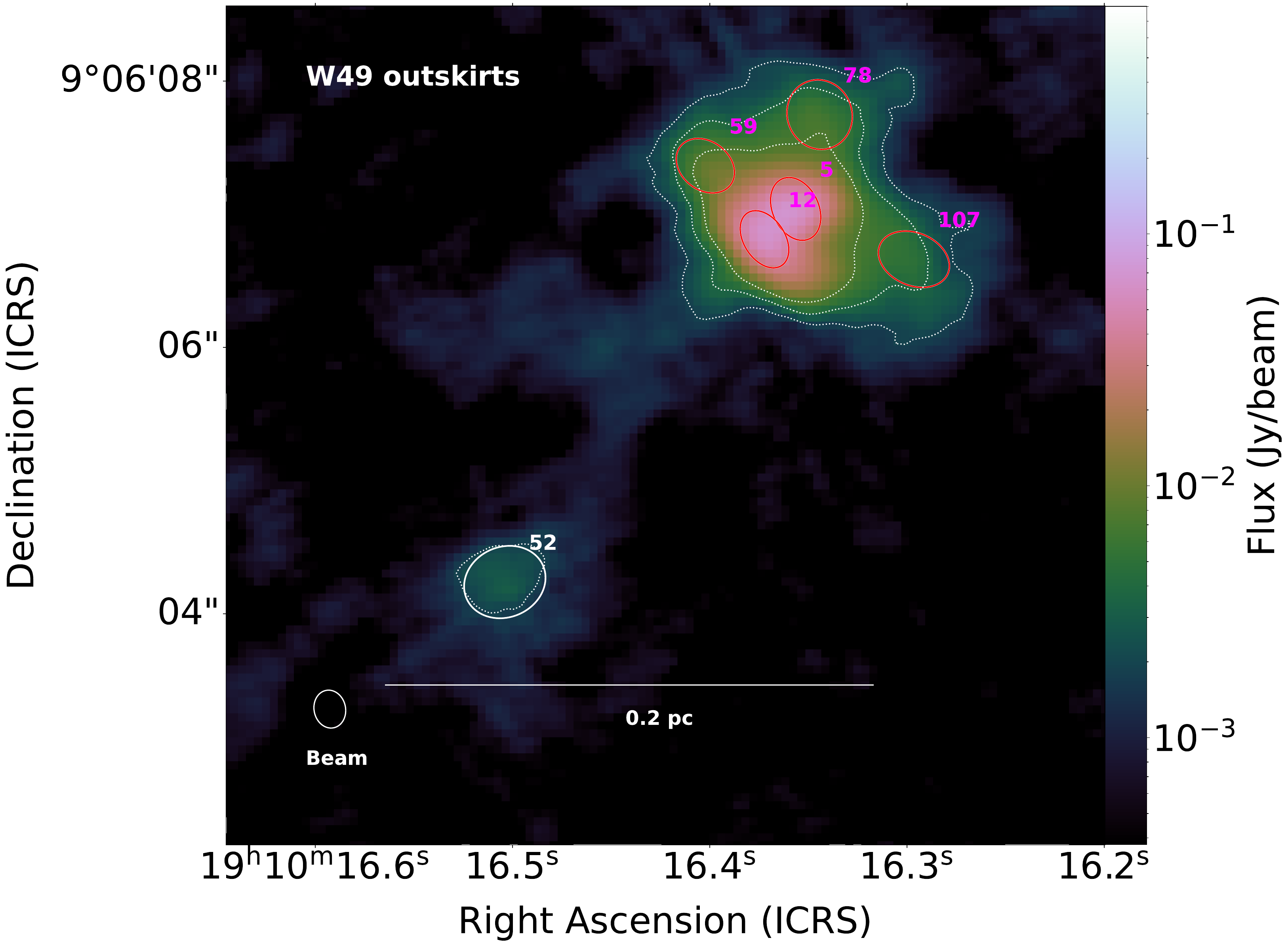}} \\
    \subfloat{\includegraphics[width=0.55\hsize]{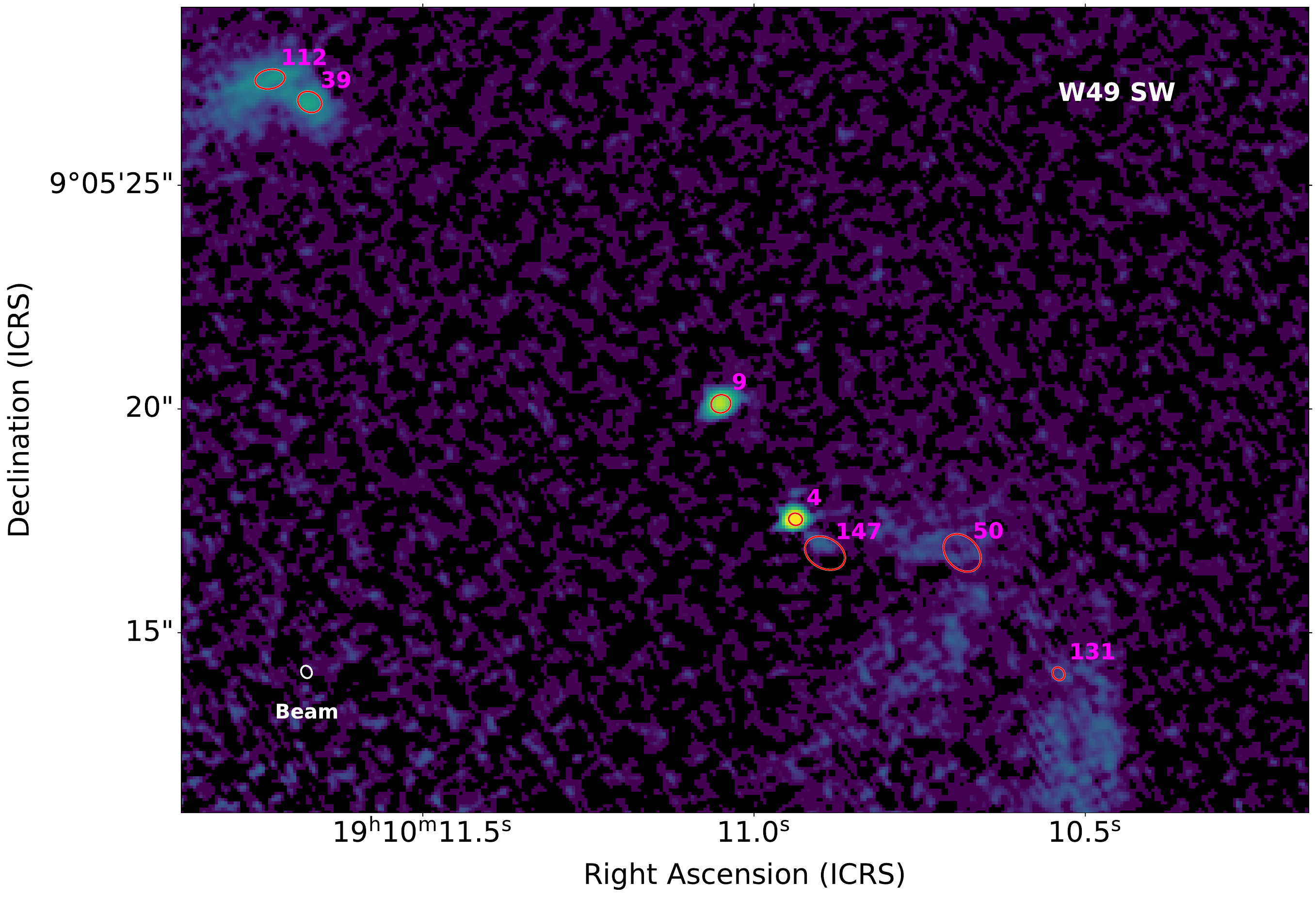}}
    \subfloat{\includegraphics[width=0.42\hsize]{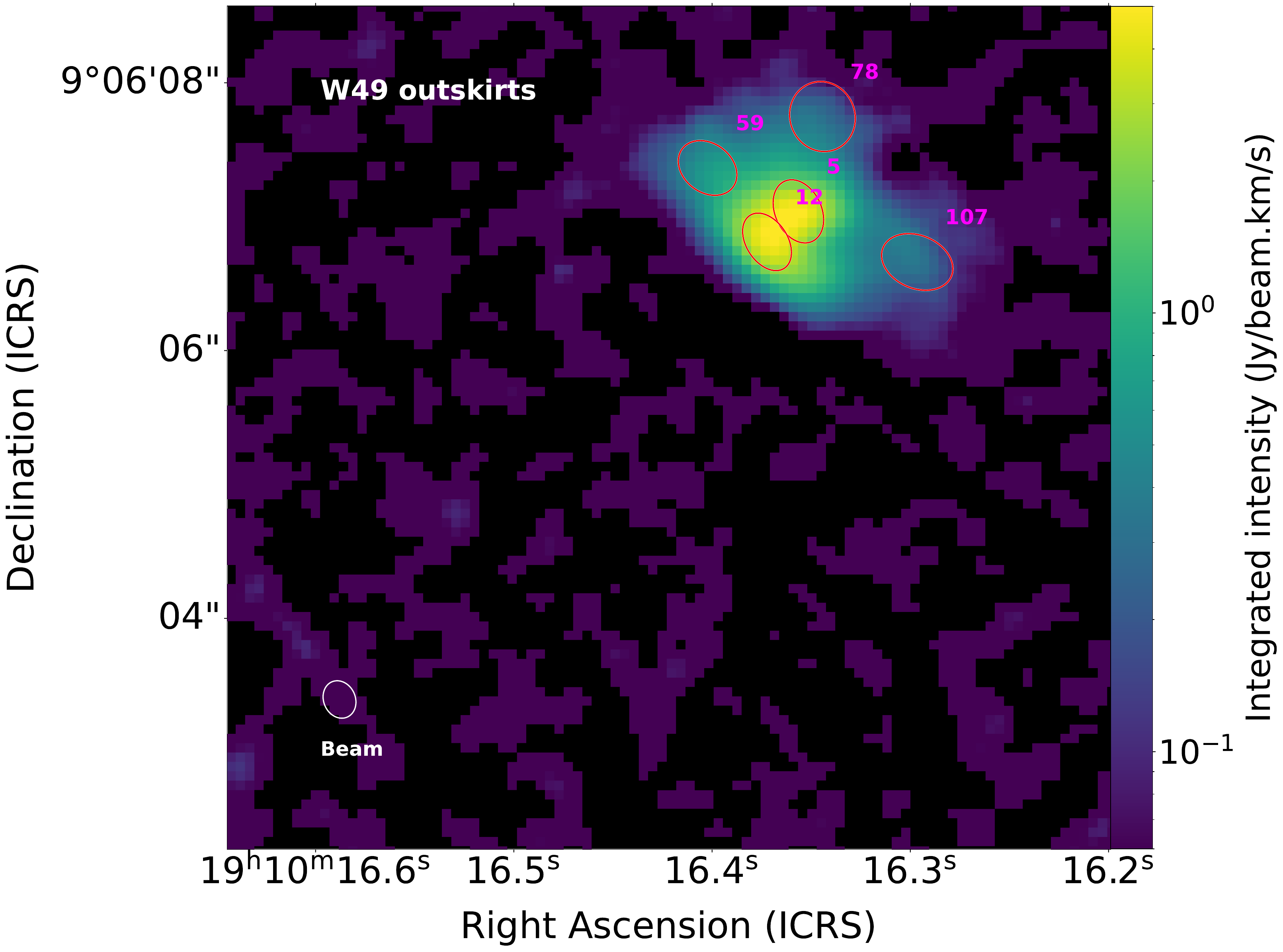}} \\
    \caption{1.3 millimeter continuum and recombination line maps toward W49 SW (left panels) and the eastern outskirts (right panels). Top panel: B6 continuum image. Contours are 5, 10, and 20 sigma. Bottom panel: moment 0 map of H30$\alpha$ line. The same conventions of \cref{fig:zomm-center} apply for ellipses and lines.}
    \label{fig:zomm-swout}
\end{figure*}

\subsection{Core catalog}
\label{su:catalog}

We used the \texttt{getsf} algorithm \citep{2021Mensh} to extract sources from the 1.3~mm continuum images. \texttt{getsf} decomposes the image into various spatial scales and separates compact sources from their background and filamentary structures. It is therefore well suited to find cores in complex environments such as high-mass star-forming regions. \texttt{getsf} has been extensively used in  ALMA studies of protocluster regions \citep[e.g.,][]{2022Pouteau,2023Nony}. 
We found 129 cores passing the post-selection filters for robust detection and measurements recommended by \citet{2021Mensh}: signal-to-noise for the background-subtracted peak larger than two, source size smaller than four times the  FWHM beamsize, and ellipticity smaller than two. Table~\ref{tab:cores-par} lists the retrieved parameters of the identified cores: position, major and minor axis FWHM, and position angle.

We also used \texttt{getsf} to measure the peak intensity and integrated flux of the cores in the ALMA B6 ($S^{\rm peak}_{\rm B6}$, $S^{\rm int}_{\rm B6}$) and VLA X-band ($S^{\rm peak}_{\rm BX}$, $S^{\rm int}_{\rm BX}$) maps. Cores are marked in the B6 continuum map in \cref{fig:contB6} and in zoom-in \cref{fig:zomm-center,fig:zomm-swout}.  
Complementary, we measured directly in the continuum maps the maximum intensity of cores in both B6 and X band, $F^{\rm max}$\footnote{Contrarily to the peak intensity in X band $S^{\rm peak}_{\rm BX}$, which is measured at the center of the ellipse defined in the B6 map, $F^{\rm max}_{\rm BX}$ is measured on the brightest pixel within the ellipse.}, and computed the integrated flux from the sum of the pixel intensities ($I$) within the core FWHM: $F^{\rm int}=c\,\Sigma I\,A_{\rm pix}/A_{\rm beam}$; with $A_{\rm pix}$ the pixel area and $A_{\rm beam}=\pi\,\theta_{min}\theta_{maj}/(4\mathrm{ln(2)})$ the beam area. We set the constant c=1/ln(2) to account for the core extension beyond its FWHM. We used this second set of parameters for the calculations presented in \cref{su:ff_B6}, in which we compare a continuum map dominated by thermal dust emission (ALMA B6 at 230~GHz) with one dominated by free-free emission (VLA X-band at 9~GHz).  

\subsection{Recombination line and free-free emission} 
\label{su:RLandVLA}
We identified cores associated with H/UC\,\textsc{Hii} regions using two tracers of ionized gas, the H30$\alpha$ recombination line and the free-free emission measured using the X-band map.

\subsubsection{Detection of the H30$\alpha$ line}
\label{sus:RL}
We identified cores associated with H30$\alpha$ emission using two complementary methods.
Our first method is as follows:  
for all the cores we inspected the spectrum at the frequency of the H30$\alpha$ line and performed Gaussian fitting to measure the line systemic velocity, the FWHM line width $\Delta V$, and the peak brightness temperature $T_\mathrm{B}$. 
We then separated cores with  brightness temperature $T_\mathrm{B} > 2$ K (38 cores) from cores with $T_\mathrm{B}$ between 0.8 and 2 K (14 cores). 
There are $^{33}$SO$_2$~(12$_{3,9}$ -- 12$_{2,10}$) lines very close in frequency (231.9004 GHz, 231.9015 GHz) to the H30$\alpha$ line at 231.9009 GHz, as well as CH$_3$C$^{15}$N and CH$_3$OCHO lines within 3~MHz. Therefore, disentangling the H30$\alpha$ emission can be challenging. Since the H30$\alpha$ line is expected to be much broader \citep[see, e.g., the review of][]{2007prpl.conf..181H}, we further applied a threshold for the linewidth of  $\Delta V>20~\kms$ to filter out cores dominated by contamination from molecular lines. This gave us 29 remaining cores with strong  H30$\alpha$ detection ($T_\mathrm{B} > 2$ K) and 9 with a weak detection.

Our second method relies on moment 0 maps of the H30$\alpha$ line, which we constructed by integrating in velocity from $-50~\kms$ to $+50~\kms$ (the line systemic velocity ranges from about 0 to 20 $\kms$ LSR). We considered a positive detection when the core's FWHM ellipse includes a group of pixels above a given threshold. Considering the large dynamic range of the line emission in the map, we noticed that the noise strongly decreases from the center to the edges of the mosaic. In consequence, we defined two reference thresholds, 5$\sigma_{\rm ctr}=0.4$ Jy\,beam$^{-1}$$\kms$ in the center, 5$\sigma_{\rm sw}=0.1$ Jy\,beam$^{-1}$$\kms$ in the South West (SW) region, and we constructed a clipped moment 0 map where pixels below $5\sigma$ are set to zeros. 
All the 29 cores with a strong H30$\alpha$ line detection are also detected in the moment 0 maps and qualify as robust H30$\alpha$ detections. 
On the other hand, the nine cores with  weak H30$\alpha$ line detection using the first method have  a few isolated pixels above their respective threshold  
at most, and are considered as tentative H30$\alpha$ detections. 
Cores with a robust or tentative H30$\alpha$ detection are overlaid on the moment 0 maps in the bottom panels of \cref{fig:zomm-center,fig:zomm-swout}.

\subsubsection{Detection of free-free centimeter emission}
\label{sus:VLA}
We defined the criterion for a robust detection and measurement in X band as $S^{\rm peak}_{\rm BX} > 3\sigma_{\rm BX}$ and $F^{\rm max}_{\rm BX} > 3\sigma_{\rm BX}$, where $\sigma_{\rm BX}=0.22$\,mJy\,beam$^{-1}$ is the noise in the X band maps convolved to the B6 angular resolution, measured using the median absolute deviation.
This criterion combines the background subtracted peak intensity $S^{\rm peak}$ determined by \textit{getsf} and the maximum intensity measured in the original continuum map (see \cref{su:catalog}).
Out of the 109 cores lying in the field of view of the X band map, 27 have a robust detection at 3.3 cm. Table~\ref{tab:cores-par} lists the integrated core fluxes in X band ($F^{\rm int}_{\rm BX}$) and Band 6 ($F^{\rm int}_{\rm B6}$). 

\subsection{Identification of hot molecular cores}
\label{su:hotcore}

We used methyl formate (CH$_3$OCHO) as a signature of hot molecular core emission, following the method described in \citet{2022Brouillet} where it was found that the use of methyl formate to identify the main HMCs in the W43-MM1 region provides similar results compared to the use of a broad frequency band with various other lines of complex organic molecules (see their Fig. 5). This study showed that this molecule offers a good compromise between widespread detectability and reliability in tracing HMC. 
Methyl formate has also been used to identity HMC candidates in the recent survey of \citet{2024Bonfand}. 
We have based the identification on the CH$_3$OCHO~(17$_{3,14}$ -- 16$_{3,13}$) doublet at 218.281 and 218.298~GHz, with an upper-level energy $E_{\rm u}$ of 99.7~K. We have further verified our detections with a synthetic spectrum fitting the numerous methyl formate lines in Spw~3. 
We found 19 dust cores with HMC emission within the entire ALMA mosaic, that is 17 within the field of view of the X band map and 2 in the south-western region. They are indicated with an asterisk in \cref{tab:cores-par}.

\vfill
\section{Characterization of dust cores}
\label{s:charac}
From the measurements presented in \cref{s:analy}, we identify cores associated with H/UC\,\textsc{Hii} regions in \cref{su:HII} and measure their properties in \cref{su:ff_B6}. The inclusion of HMCs enables us to refine the evolutionary status of cores in \cref{su:evol}.

\subsection{Classification of cores with H/UC\,\textsc{Hii} regions}
\label{su:HII}

\subsubsection{With VLA measurements}
For 109 cores within the field of view of the X band image, we have information on both the X band and the H30$\alpha$ detection. The 27 cores with a robust detection in X band are likely associated with an H/UC\,\textsc{Hii} region. 20 are also detected in H30$\alpha$ and three have a tentative H30$\alpha$ detection (\#74, \#87, \#116). 
The remaining four cores without detection in H30$\alpha$ (\#8, \#18, \#104, \#117) only have a weak X band detection, with $S^{\rm peak}_{\rm BX}$ and $F^{\rm max}_{\rm BX}$ between 3 and 10$\sigma_{\rm BX}$, which suggests that our sensitivity to detect ionized gas is higher in the X band map than with the recombination line.   
Two other cores have a tentative H30$\alpha$ detection and no detection in X band: after visual inspection core \#94 is considered as an H/UC\,\textsc{Hii} region candidate and core \#82 is rejected.  

Our identification of dust cores associated with H/UC\,\textsc{Hii} regions largely overlaps with the independent dendrogram catalog of H/UC\,\textsc{Hii} regions of Juárez-Gama et al. (in prep.) (see \cref{su:VLAdat}). 21 H/UC\,\textsc{Hii} dendrogram structures out of 79 within the X band map are associated with a core with H/UC\,\textsc{Hii} region, as previously defined. 
In detail, 20 \textsc{Hii} dendrogram structures are associated with 26 cores with a robust detection in X band\footnote{5 dendrogram \textsc{Hii} structures are associated with more than one dust core.} 
Only one \textsc{Hii} dendrogram structure is associated with a dust core (\#27) not detected in X band.  Conversely, only one dust core with a detection in X band is not associated with a dendrogram structure. This core (\#18) is among the weakest X band detections  ($S^{\rm peak}_{\rm BX} = 4\sigma_{\rm BX}$ and $F^{\rm max}_{\rm BX} = 3.4\sigma_{\rm BX}$) and is not found in H30$\alpha$. 
Fifity seven structures in the dendrogram catalog built from the X band map do not have any counterpart in the Band 6 core catalog and qualify as H/UC\,\textsc{Hii} regions without association to dust cores. A summary of the relation  between the 28 dust cores associated with H/UC\,\textsc{Hii} regions (27 when excluding the H/UC\,\textsc{Hii} region candidate) and the dendrogram-identified H/UC\,\textsc{Hii} regions is shown in \cref{fig:statbox}. 

\subsubsection{Without VLA measurements}
For the 20 dust cores outside the field of view of the X band image, the recombination line is the only tool to search for emission from H/UC\,\textsc{Hii} regions. Nine cores associated with a robust H30$\alpha$ detection (four in the southwestern region, five in the eastern outskirts, see \cref{fig:statbox}) are likely associated with an H/UC\,\textsc{Hii} region. 
After visual inspection, three additional cores with a tentative H30$\alpha$ detection located in the SW region (\#50, \#131, \#147) are considered as H/UC\,\textsc{Hii} region candidates, and another core located in the eastern outskirts (\#115) is rejected.

\subsection{Free-free emission in ALMA Band 6}
\label{su:ff_B6}
\begin{table}[ht] 
\caption{Derived parameters of dust cores associated with H/UC\,HII regions: core identifier, integrated free-free flux at 1.3 mm estimated from the 3.3\,cm map and from the H30$\alpha$ line, percentage of free-free emission in Band 6, spectral index of free-free emission, and emission measure.} 
\label{tab:HII-par} 
\begin{tabular}{ccc|ccc} 
\hline 
n\tnote{a} & $F^{\rm int}_\mathrm{ff,cm}$\,\tnote{c} & $F^{\rm int}_\mathrm{ff,RL}$\,\tnote{c} & $P_\mathrm{ff}$ & $\alpha$ & EM \\ 
  & [mJy.b$^{-1}$] & [mJy.b$^{-1}$] & [\%] & & [pc\,cm$^{-6}$] \\ 
\hline 
  1 & 26.3 & 809.1 & >78 & >0.96 & >2.9e8 \\ 
  2 & 27.6 & 422.5 & >58 & >0.74 & >2.9e8 \\ 
  3 & 40.3 & 224.2 & >61 & >0.43 & >2.9e8 \\ 
  4 & - & 90.4 & 96 & - & - \\ 
  5 & - & 71.5 & 54 & - & - \\ 
  7 & 29.2 & 115.3 & >61 & >0.32 & >2.9e8 \\ 
  8 & 1.1 & - & - & - & 1.1e7 \\ 
  9 & - & 55.8 & 65 & - & - \\ 
 10 & 8.8 & 179.6 & 54 & 0.83 & 1.8e8 \\ 
 11 & 15.8 & 24.6 & 44 & 0.04 & 1.4e8 \\ 
 12 & - & 65.3 & 56 & - & - \\ 
 13 & 9.2 & 94.8 & 64 & 0.62 & 1.1e8 \\ 
 14 & 97.1 & 137.1 & >67 & >0.00 & >2.9e8 \\ 
 15 & 43.2 & 39.5 & 67 & -0.13 & 1.9e8 \\ 
 16 & 49.6 & 176.5 & 63 & 0.29 & 2.9e8 \\ 
 17 & 12.7 & 80.5 & 64 & 0.47 & 1.7e8 \\ 
 18 & 0.7 & - & - & - & 4.5e6 \\ 
 23 & 9.1 & 11.0 & 62 & -0.04 & 8.3e7 \\ 
 29 & 12.4 & 17.6 & 75 & 0.01 & 9.5e7 \\ 
 39 & - & 17.9 & 59 & - & - \\ 
 40 & 39.4 & 40.5 & 51 & -0.09 & 1.5e8 \\ 
 44 & 1.0 & 8.1 & 37 & 0.53 & 8.7e6 \\ 
 46 & 24.9 & 35.3 & 61 & 0.00 & 3.1e8 \\ 
 47 & 14.6 & 23.7 & 54 & 0.05 & 1.6e8 \\ 
 50 & - & 9.1 & 25 & - & - \\ 
 59 & - & 11.2 & 49 & - & - \\ 
 69 & 12.4 & 13.6 & 64 & -0.07 & 4.2e7 \\ 
 72 & 22.4 & 19.1 & 45 & -0.15 & 1.1e8 \\ 
 73 & 21.6 & 25.0 & 55 & -0.06 & 1.7e8 \\ 
 74 & 2.6 & 2.9 & 33 & -0.07 & 1.4e7 \\ 
 78 & - & 8.1 & 37 & - & - \\ 
 87 & 10.0 & 13.7 & 42 & -0.01 & 1.3e7 \\ 
 94 & - & 13.8 & 54 & - & - \\ 
104 & 0.5 & - & - & - & 9.1e6 \\ 
107 & - & 7.1 & 48 & - & - \\ 
112 & - & 16.7 & 65 & - & - \\ 
116 & 2.4 & 3.6 & 73 & 0.02 & 9.9e6 \\ 
117 & 1.7 & - & - & - & 1.3e7 \\ 
131 & - & 1.3 & 40 & - & - \\ 
147 & - & 4.4 & 22 & - & - \\ 
\hline 
\end{tabular} 
\end{table} 

Out of the 129 dust cores detected with ALMA over the entire mosaic, we found 36 cores associated with H/UC\,\textsc{Hii} regions (27 with VLA measurements, 32 with H30$\alpha$ measurements, see \cref{su:HII}) and 4 cores tentatively associated with H/UC\,\textsc{Hii} regions  (\#50, \#94, \#131, \#147), which have only tentative H30$\alpha$ detections.

A first evaluation of their free-free emission at 1.3 mm  is obtained from the integrated fluxes, $F^{\rm int}_{\rm BX}$ measured in the X band map over FWHM of the cores (see \cref{su:catalog}):
\begin{equation}
F^{\rm int}_\mathrm{ff,cm} = F^{\rm int}_\mathrm{BX}\,\left(\frac{\nu_\mathrm{B6}}{\nu_\mathrm{BX}} \right) ^{\alpha},  
\end{equation}

\noindent 
where $\nu_\mathrm{B6}=227.95$~GHz and $\nu_\mathrm{BX}=9.11$~GHz are the reference frequency of the ALMA B6 and VLA X band images, respectively (see \cref{tab:data}).
We took a spectral index $\alpha=-0.1$, assuming that the free-free emission is optically thin all the way from 3 cm to 1 mm. The fluxes calculated in this way are thus lower limits to the free-free emission in the ALMA image. 

\newpage
We also constructed a pixel by pixel free-free map $I_\mathrm{ff,RL}$ using the moment 0 of the H30$\alpha$ recombination line \citep[see, e.g.,][]{Wilson09,2019ApJ...871..185L}: 
\begin{equation}
    \centering
    I_\mathrm{ff,RL} = 1.432 \times 10^{-4} \bigl( \nu_0^{-1.1} T_\mathrm{e}^{1.15} \bigr ) \bigl( 1+N_\mathrm{He}/N_\mathrm{H} \bigr ) \int I_{{\rm H}30} dv, 
\label{eq:ff_map}
\end{equation}
where the electron temperature is $T_\mathrm{e} = 7000$ K (e.g. \citealt{2023MNRAS.520.3245Z}), the helium to hydrogen number ratio is $N_\mathrm{He}/N_\mathrm{H} = 0.08$, and $\nu_0$ is the central frequency of the H$30\alpha$ line.
We then obtained a second evaluation of the free-free emission in B6 by measuring on the $I_\mathrm{ff,RL}$ map the integrated fluxes of the cores, $F^{\rm int}_\mathrm{ff,RL}$, using the method presented in \cref{su:catalog}. 

From the previous measurements, we derived various parameters for the cores associated with H/UC\,\textsc{Hii} regions. 
We defined their percentage of free-free emission in B6 as $P_\mathrm{ff}=\,F_\mathrm{ff,RL}/F_\mathrm{B6}$. $P_\mathrm{ff}$ ranges from 22 to 96\%. The dust flux in B6 is expressed as $F_\mathrm{B6, dust} = F_\mathrm{B6} - F_\mathrm{ff}$, with $F_\mathrm{ff}=F_\mathrm{ff,RL}$ when the H30$\alpha$ is detected, and $F_\mathrm{ff}=F_\mathrm{ff,cm}$ otherwise.  
By construction, $F_\mathrm{ff,RL}$ can be expressed as a function of $F_\mathrm{B6, dust}$ and $P_\mathrm{ff}$: $F_\mathrm{ff,RL} = F_\mathrm{B6, dust}\,P_\mathrm{ff}/(1-P_\mathrm{ff})$. 
We also computed the spectral index of free-free emission from $\nu_\mathrm{BX}$ to $\nu_\mathrm{B6}$: 
\begin{equation}
\alpha_\mathrm{ff} = \frac{\mathrm{log}(F_\mathrm{ff,RL}/F_\mathrm{BX})}{\mathrm{log}(\nu_\mathrm{B6}/\nu_\mathrm{BX})}
\end{equation}

Finally, we computed the emission measure, EM, of H/UC\,\textsc{Hii} emission  within dust cores: 

\begin{equation} \label{eq:EM}
\mathrm{EM}~[\mathrm{pc\,cm}^{-6}] = \int n_e^2 dl = 12\,\tau_\mathrm{ff} \Bigl[ \frac{\nu_\mathrm{BX}}{\mathrm{GHz}} \Bigr ]^{2.1}
\Bigl [ \frac{T_\mathrm{e}}{\mathrm{K}} \Bigl ]^{1.35},
\end{equation}

\noindent
with $n_\mathrm{e}$ the electron density and $\tau_\mathrm{ff}$ the free-free continuum optical depth, expressed as 
\begin{equation} \label{eq:tauff}
\tau_\mathrm{ff} = - \ln \biggl [ \biggl ( 1 - \frac{T_\mathrm{B}}{T_\mathrm{e}} \biggr ) \biggr ],
\end{equation}

The brightness temperature $T_\mathrm{B}$ is calculated from the maximum intensity in X band within the cores FWHM, $F^{\rm max}_{\rm BX}$, and we used a constant electron temperature $T_\mathrm{e} = 7000$\,K. 
Two cores with $T_\mathrm{B} > 7000$\,K (\#3 and \#16) and 4 cores with $6700 < T_\mathrm{B} < 7000$\,K (\#1, \#2, \#7 and \#14) are likely to have higher electron temperature, considering the condition $T_\mathrm{e} > T_\mathrm{B}$ set by \cref{eq:tauff}. We computed for these six cores a lower limit of the emission measure using an opacity $\tau_\mathrm{ff}=1.5$ in line with the rest of the core sample. 
The emission measures of cores with H/UC\,\textsc{Hii} regions range from 4.5$\,\times\,10^6$ to 3.1$\,\times\,10^8$ pc\,cm$^{-6}$, which are in the range of values found for ultra-compact \textsc{Hii} regions \citep{2002ARA&A..40...27C,2007prpl.conf..181H}. 
Table~\ref{tab:HII-par} lists $F_\mathrm{ff,cm}$, $F_\mathrm{ff,RL}$, $P_\mathrm{ff}$, $\alpha_\mathrm{ff}$, and EM for the 40 cores with robust and candidate H/UC\,\textsc{Hii} emission. 

\subsection{Core classification}
\label{su:evol}

\begin{figure}
    \centering
    \includegraphics[width=\hsize]{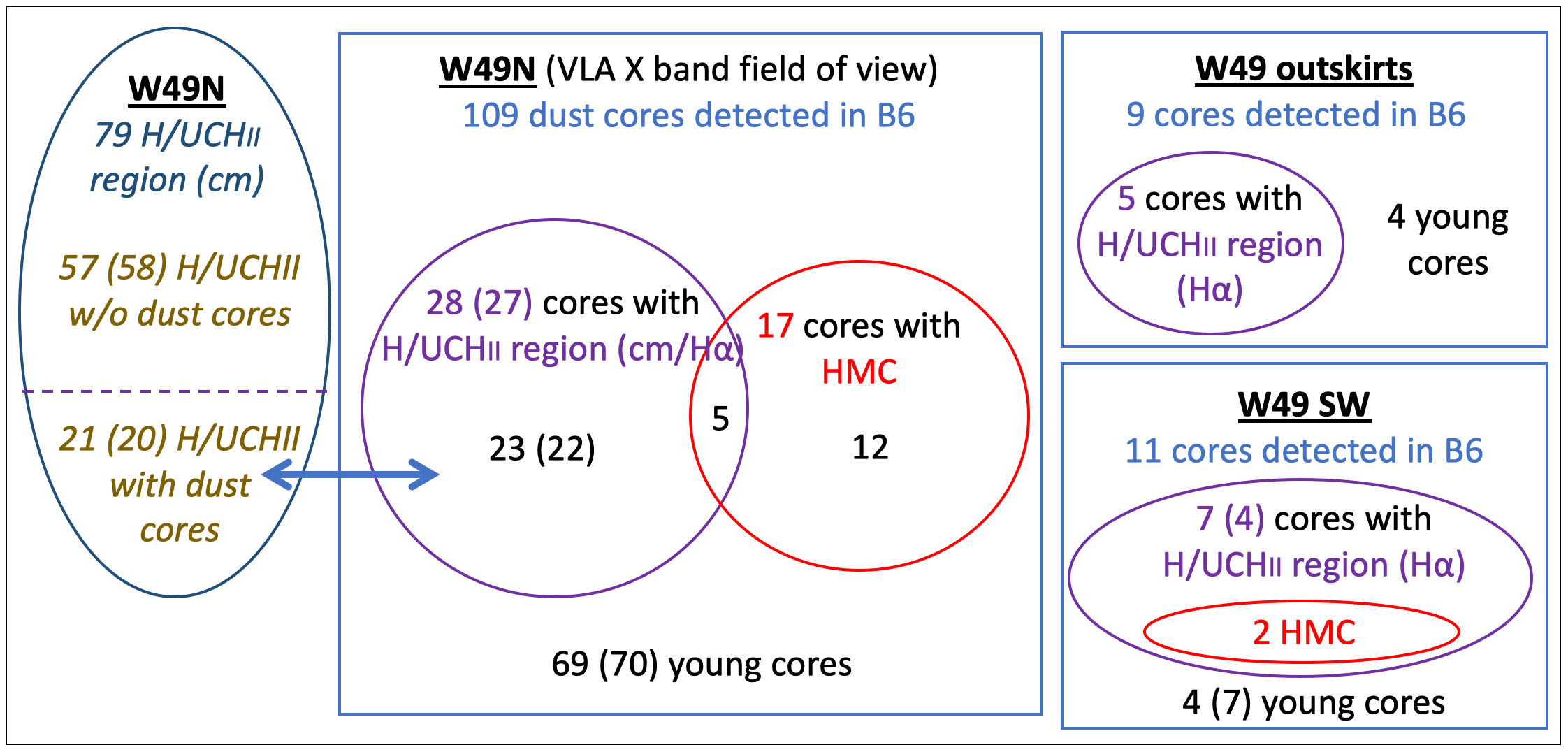}
    \caption{Classification of the dust cores extracted in the ALMA Band 6 maps, based on the detection of HMCs and/or \textsc{Hii} region emission. Cores lying within the field of view of the VLA map (W49N) are separated from cores in W49SW and in the eastern outskirts.}
    \label{fig:statbox}
\end{figure}

In what follows, we combine the statistics obtained from the detection rates of H/UC\,\textsc{Hii} regions (see \cref{su:HII}) with the hot molecular core (HMC) detections (see \cref{su:hotcore}) to build a classification of the dust cores into four types: $a/$ cores without either HMC or H/UC\,\textsc{Hii} region, $b/$ cores associated with a HMC but without H/UC\,\textsc{Hii} emission, $c/$ cores associated with both a HMC and H/UC\,\textsc{Hii} emission, and $d/$ cores associated with a H/UC\,\textsc{Hii} region but without HMC emission. 
Within the full ALMA B6 mosaic, the number counts of each type are  $a/$: 77 (81), $b/$: 12, $c/$: 7, and $d/$: 33 (29). Numbers in parenthesis correspond to the situation where all the cores with H/UC\,\textsc{Hii} regions candidates are accounted in first type.  
Within the smaller field of view of the VLA X band map, the statistics are 69 (70), 12, 5 and 23 (22). \cref{fig:statbox} shows a diagram summarizing the detection statistics.  
In addition, 57 (58) H/UC\,\textsc{Hii} regions without cores have been detected by Juárez-Gama et al. (in prep.). 

\begin{figure*}
    \centering
    \includegraphics[width=\hsize]{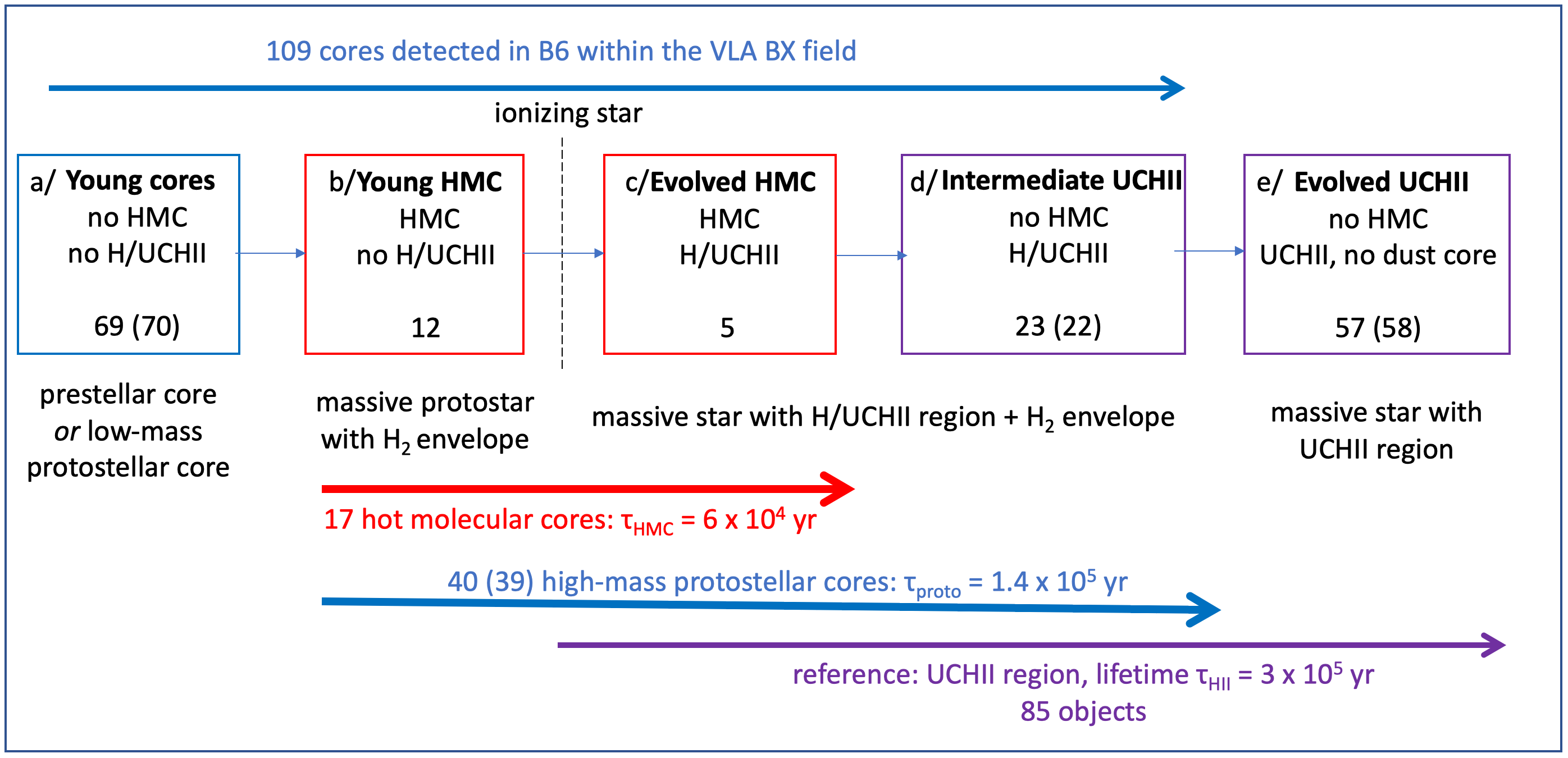}
    \caption{Evolutionary sequence established for the cores extracted from the ALMA B6 maps within the X-band field of view (W49N),  based on the presence or absence of hot molecular cores (HMC) and hyper/ultra-compact \textsc{Hii} regions (H/UCHII). Statistical lifetimes for the various evolutionary stages are computed assuming a lifetime of the H/UC\,\textsc{Hii} stage of $3\times10^5$~yr (see \cref{su:lifetime}).}
    \label{fig:timeseq}
\end{figure*}

\section{Discussion}
\label{s:discu}

\subsection{Evolution of Star Forming Cores in W49A}
\label{su:sf}
From the empirical classification presented in \cref{su:evol}, in this section we propose 
an evolutionary sequence for the dust cores in W49A, based on the presence or absence of the analyzed indicators of high-mass star formation. \cref{fig:timeseq} summarizes the proposed evolutionary scheme. 

Cores in stage $a/$ do not have evidence of a HMC or H/UC \textsc{Hii} region, they are likely prestellar or low-mass protostellar cores. 
Among cores in stage $a/$, which we label "young cores", only a fraction are the precursors of high-mass stars. It is indeed expected that some of them will produce low-mass objects. 
From stage $b/$ onward, all cores are expected to harbor a high-mass ($\gtrsim 8~\Msol$) protostar.  As it grows in mass, the protostellar object will first reach the luminosity necessary to heat its envelope to a temperature where complex organic molecules are released from dust grains. This defines the onset of the hot molecular core stage ($b/$, "young HMC" in \cref{fig:timeseq}). Later on, when the extreme ultraviolet (EUV) radiation of the star is sufficient to ionize atomic hydrogen, an \textsc{Hii} region is born. During the short phase in which the HMC and the H/UC\,\textsc{Hii} region coexist (stage $c/$, "evolved HMC"), we expect that the remaining dust and molecules form a cocoon around the \textsc{Hii} region, which is mostly free of dust \citep{2008ASPC..387..232L}.  
The radiation then becomes strong enough to fully destroy molecules, and only some dust remains surrounding the H/UC\,\textsc{Hii} region (stage $d/$, "intermediate UC\textsc{Hii}"). Finally, the UC\textsc{Hii} fully clears its way out of the now extinct core, which is not detected anymore in dust emission 
(stage $e/$, "evolved UC\textsc{Hii}"). 
We expect that high-mass protostars keep accreting from their envelopes from stages $b/$ to $d/$, which therefore represent the phases of high-mass protostellar evolution. Under this definition, 40 (39) cores out of 109 in W49N are high-mass protostellar cores. 
For W49SW and what we label as outskirts,  12 (9) cores out of 20 contain high-mass protostars (see \cref{fig:statbox}).

Regarding their spatial distribution, the large majority of the cores detected at 1.3\,mm (109 out of 129) lie within the central star forming region W49N. Only a few cores are located at distances beyond a parsec from the central hub (see \cref{fig:contB6}).
In W49N, HMCs and the well known \textsc{Hii} regions \citep{1997ApJ...482..307D,2018ApJ...863L...9D} co-exist with a population of young cores (see zoom-in \cref{fig:zomm-center}). This indicates that the central protocluster is still actively forming stars in spite of the significant stellar feedback, as suggested by  \citet{2013ApJ...779..121G} from the measurement of a $\sim 2\times10^5~\Msol$ molecular gas reservoir in W49N. A similar scenario has been proposed for the also very active W51-E and W51-IRS2 protoclusters \citep{2016A&A...595A..27G}. 

Using uncertainties from the binomial distribution, we measured the 75\% confidence interval for the percentage of cores containing high-mass protostars (stages $b/$ to $d/$) to be [30-42]\% in W49N and [30-75]\% in W49SW and the W49 outskirts. Therefore, the relative abundance of high-mass protostars does not appear to be statistically different between the center of W49A and its periphery. 
We reach the same conclusion when only HMCs or cores with H/UC\,\textsc{Hii} regions are considered.
This suggests that the different sub-protoclusters analyzed in this paper have a similar "age", in the sense that their core and protostellar populations look statistically similar. The small number of objects in W49SW and the W49 outskirts forbids us to make a stronger statement, but this further suggests that star formation across the W49A cloud started simultaneously, possibly by cloud-cloud collision as suggested by \citet{2022PASJ...74..128M}. 

\subsection{Ionization evolution within dust cores}
\label{su:ioni}

\begin{figure*}
    \centering
    \includegraphics[width=0.49\hsize]{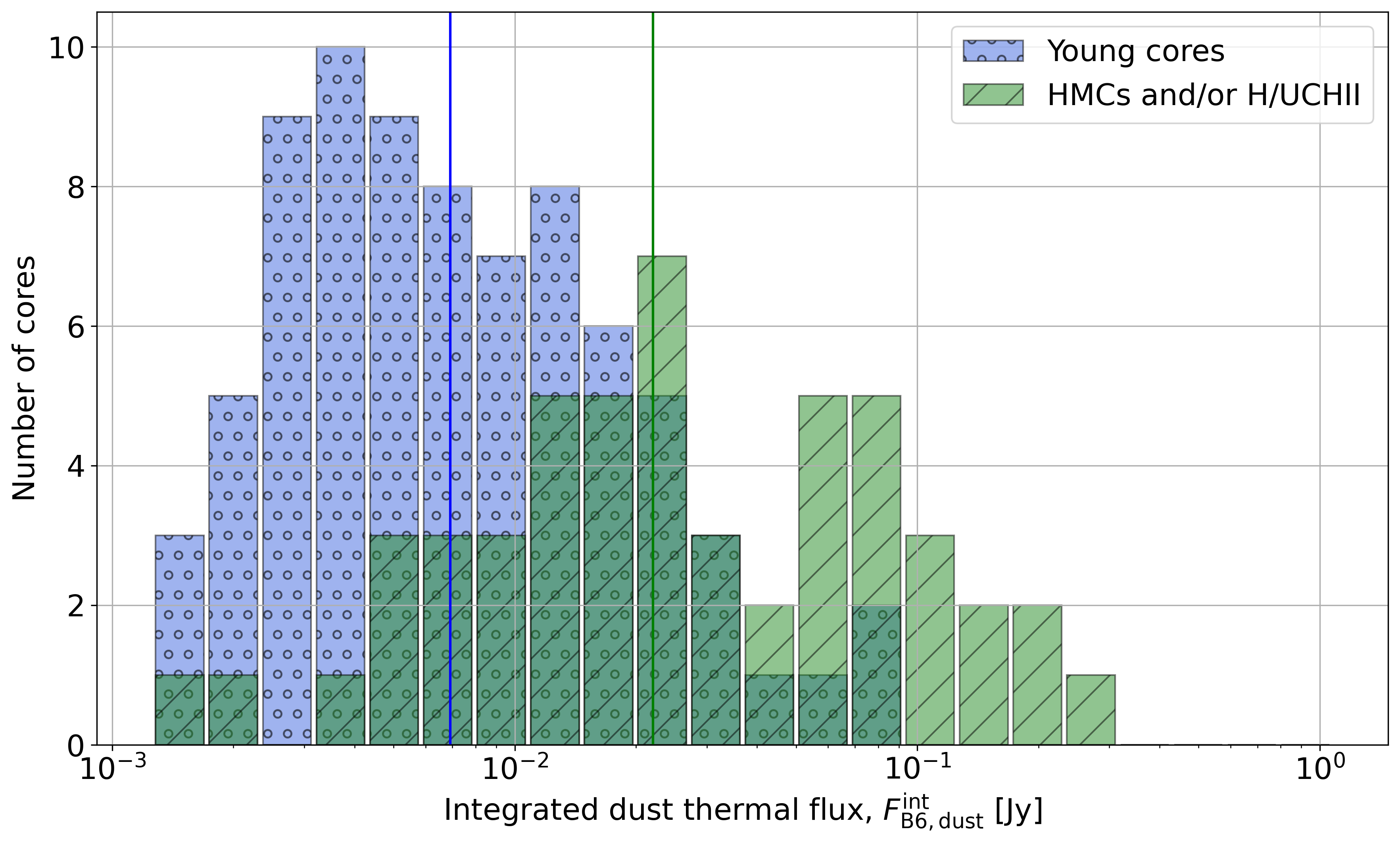}
    \includegraphics[width=0.49\hsize]{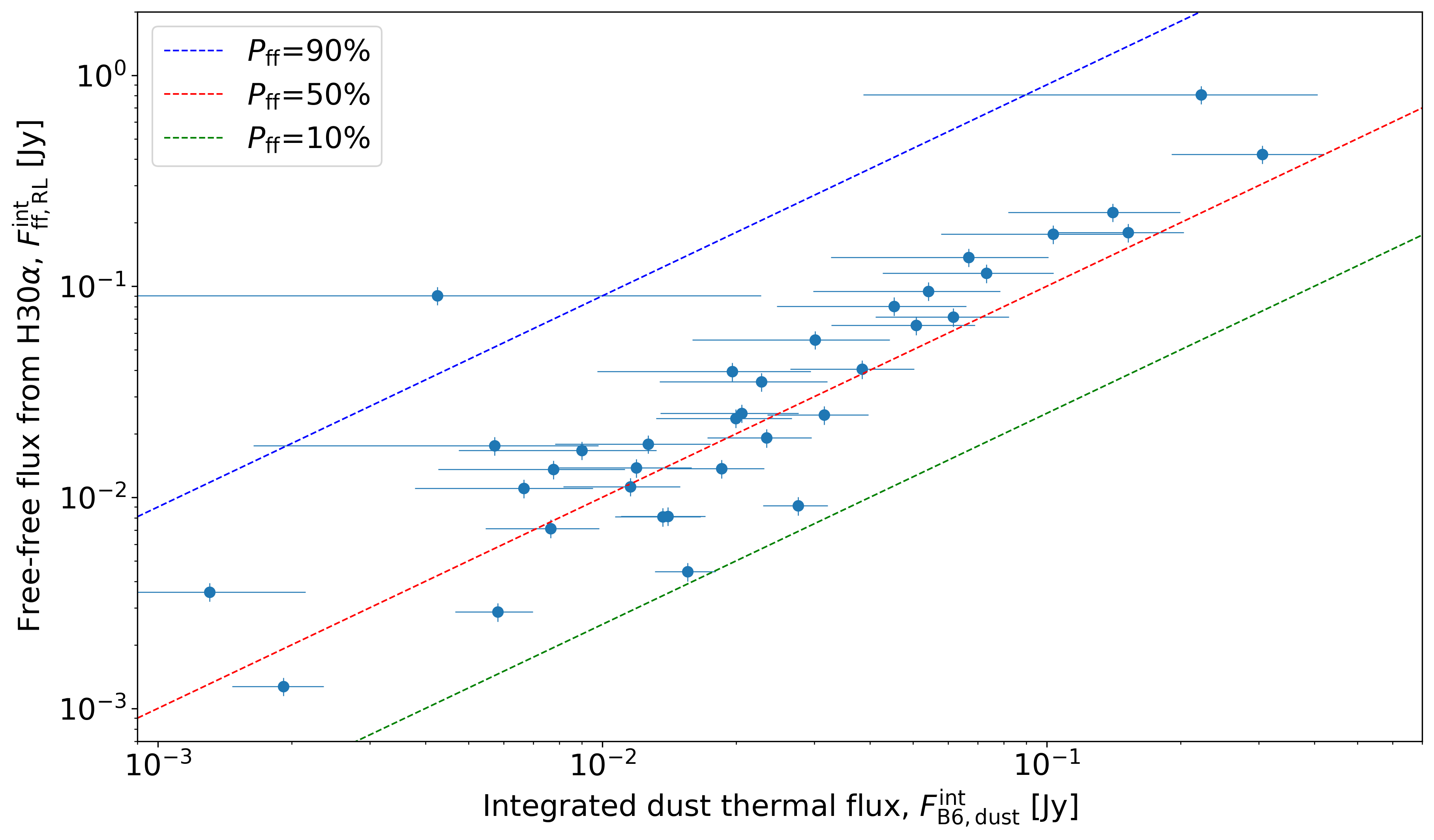}
    \caption{ 
    Left: Distribution of the core-integrated dust fluxes. Cores with hot molecular core emission and/or H/UC\,\textsc{Hii} region (stages $b/$ to $d/$, shown in green) are compared to the "young cores" without evidence of massive star formation (stage $a/$, shown in blue). Vertical lines represent the median value of the distributions. 
    Right: Comparison between the free-free and dust fluxes. The core sample in this plot is made of the 35 cores with (robust or tentative) H$30\alpha$ detection, classified as cores with H/UC\,\textsc{Hii} region. Dashed lines represent constant values of the percentage of free-free emission $P_\mathrm{ff}$ with respect to the total core flux. Uncertainties are evaluated as 10\% of the free-free and B6 fluxes.
    }
    \label{fig:compar-plots1}
\end{figure*}

\begin{figure}
    \centering
    \includegraphics[width=0.98\hsize]{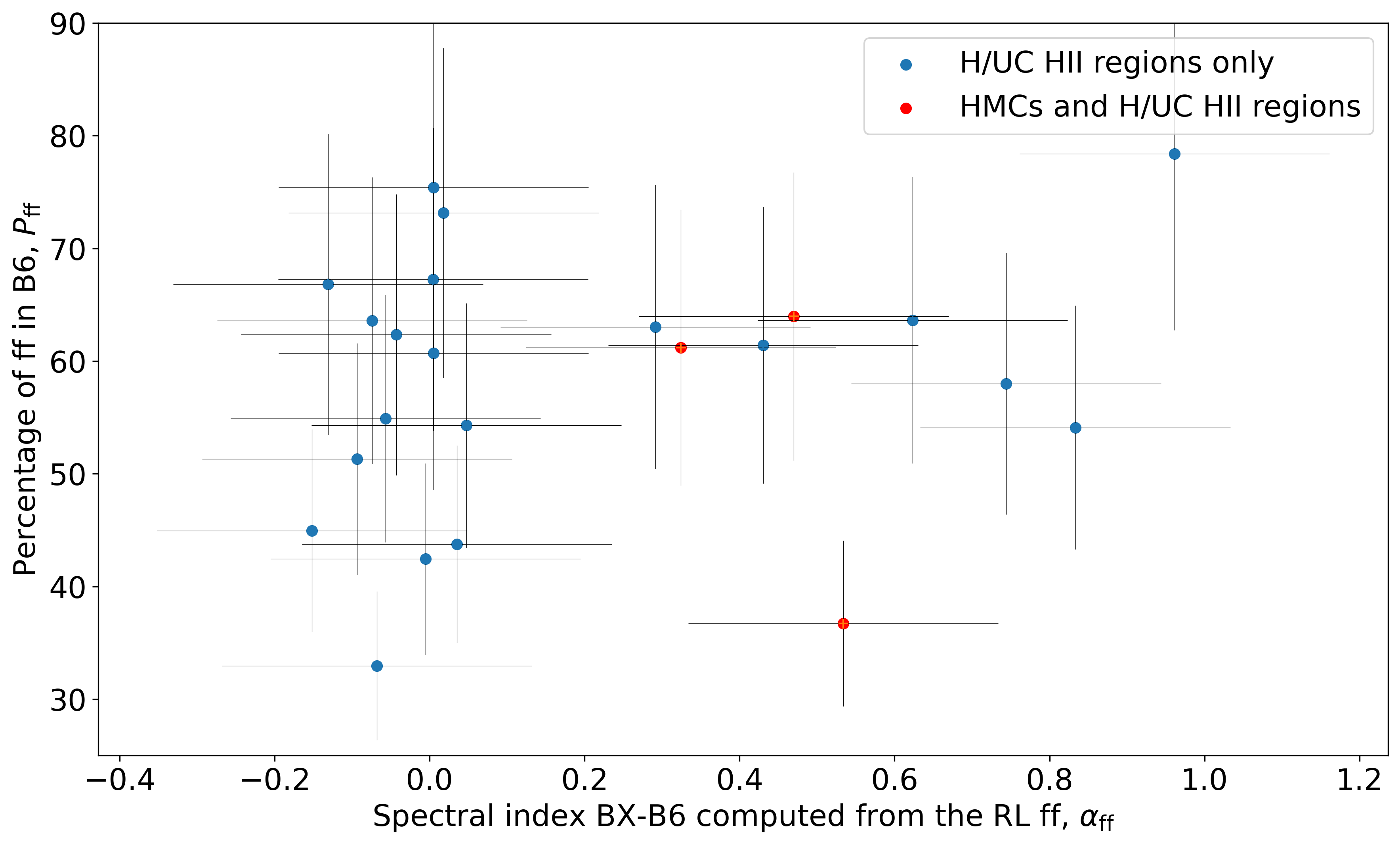}
    \caption{The percentage of free-free emission $P_\mathrm{ff}$ at 1.3 mm (y-axis) is not anti-correlated with the spectral index of free-free emission (x-axis), from which two populations of cores can be distinguished. The core sample represented is 23 cores with H$30\alpha$ and X band detections. Cores with H/UC\,\textsc{Hii} regions and HMCs are show in red, cores with H/UC\,\textsc{Hii} regions only are shown in blue.}
    \label{fig:compar-plots2}
\end{figure}

The physical parameters derived in \cref{su:ff_B6} allow us to evaluate the evolution of the protostar and envelope masses. 
The left panel of \cref{fig:compar-plots1} shows that cores with a HMC and/or H/UC\,\textsc{Hii} region have statistically higher dust flux than the others. The median dust flux of the two distributions are $2.2 \times 10^{-2}$ Jy and $6.9 \times 10^{-3}$ Jy, respectively. We have verified with a KS test that the distributions are statistically distinct ($p$-value of $4 \times 10^{-7}$). 
Since the dust flux is related to the molecular gas mass of the core, this difference suggests that the more evolved cores could also be more massive. 
Complementary, the right panel of \cref{fig:compar-plots1} shows a strong correlation between the dust flux at 1.3 mm and the amount of free-free emission. Both the Pearson and Spearman correlation coefficients are higher than 0.8, with $p$-values $\simeq 10^{-10}$. 
Since the amount of free-free emission is related to the mass of the central star through the number of ionizing photons \citep[e.g.,][]{1994ApJS...91..659K}, this correlation could indicate that the mass of the central (proto)star grows simultaneously with the mass of its gas envelope. This hypothesis is in line with models of continuous accretion across multiple scales \citep[e.g.,][]{2009MNRAS.400.1775S,2019MNRAS.490.3061V}, as well as with recent observations suggesting that cores grow in mass as they evolve from quiescent (candidate prestellar) to active protostellar stages \citep{2023Nony}. 

However, the dust flux is also dependent on the dust temperature, which might be higher in the more evolved cores than in the younger ones. This could  partially explain the difference between the distributions in the left panel of \cref{fig:compar-plots1}. 
For an envelope internally heated by an embedded protostar, the dust temperature depends on its luminosity \citep{1976ApJ...206..718S}, hence on the protostellar mass. Therefore, the correlation found in the right panel of \cref{fig:compar-plots1} between the dust and free-free fluxes could alternatively be explained by the dependence on the stellar mass of both quantities. Detailed measurements of core temperatures, which are beyond the scope of this work, are thus needed to draw firmer conclusions.

In \cref{fig:compar-plots2} we compare the percentage of free-free emission in the 1.3 mm flux ($P_\mathrm{ff}$) to the free-free spectral index ($\alpha_\mathrm{ff}$) between 3.3\,cm and 1.3\,mm, for the sample of dust cores with an estimation of their free-free content from both a centimeter continuum and H$30\alpha$ detection (see \cref{su:ff_B6}). 
A first group of dust cores with optically thin  free-free emission ($\alpha_\mathrm{ff} \simeq-0.1$ to 0) has $P_\mathrm{ff}$ spanning a large range of values, from 30 to 80\%. In contrast, a second group of cores with  partially optically thick free-free emission ($\alpha_\mathrm{ff} \sim$ 0.3 to 1) tends to show a more restricted  $P_\mathrm{ff} \simeq 55-65\%$, with a couple of exceptions. 
To first order, \textsc{Hii} regions are expected to smoothly evolve as they ionize and break out of their parental core \citep{2007ApJ...666..976K, 2016ApJ...818...52T}. During this process, they are expected to transition 
from denser objects with partially optically thick emission, to being less dense and optically thin \citep[e.g.,][]{2008ApJ...672..423K, 2009ApJ...706.1036G}. HMCs represent an early stage within this sequence (see \cref{su:sf}), therefore the fact that the three HMCs within the sample of dust cores with \textsc{Hii} emission have partially optically thick free-free emission is consistent with this scenario. Under this simple interpretation, however, the points in \cref{fig:compar-plots2} should follow an anti-correlation, or at least the -- more evolved -- optically thin \textsc{Hii} regions should have on average a larger $P_\mathrm{ff}$ than the -- younger -- optically thicker ones. This is not observed.  Previous work on hydrodynamical simulations with ionization feedback and synthetic observations has shown that the stochastic interactions of the neutral gas and ionizing photons can cause variations in the appearance of the youngest \textsc{Hii} regions \citep{2010ApJ...711.1017P}, including their spectral indices \citep{2010ApJ...719..831P}. 
This means that, even if H/UC\,\textsc{Hii}s eventually break free from their cores on timescales of $\sim 10^5$ yr, their early evolution is not necessarily monotonic. We propose that this could be the cause of the observed spread of values in \cref{fig:compar-plots2}.


\subsection{Lifetimes}
\label{su:lifetime}

We now proceed to evaluate the core lifetimes  
using the evolutionary sequence presented in \cref{su:sf} and shown in \cref{fig:timeseq}. For this purpose, we consider the population of 109 cores in W49N, for which the more complete search of H/UC\,\textsc{Hii} regions has been made.
Summing up 28 cores with H/UC\,\textsc{Hii} regions and 57 H/UC\,\textsc{Hii} regions not associated to a dust core, 85 objects fall in the H/UC\,\textsc{Hii} regions category. 
Under the hypothesis of constant star-formation and assuming that HMCs are precursors of H/UC\,\textsc{Hii} regions, the relative number of HMCs to H/UC\,\textsc{Hii} regions (17/85) provides an estimation of their statistical lifetime. Using the typical lifetime for H/UC\,\textsc{Hii} regions of $3\times10^5$~yr, calculated from Galaxy-wide surveys  \citep{2002ARA&A..40...27C,2011ApJ...730L..33M}, we estimate the lifetime of the HMC phase to be $\tau_\mathrm{HMC} \simeq (17/85) \cdot (3 \times 10^5) = 6\times10^4$\,yr. 
The ratio of 17 HMCs to 85 H/UC\,\textsc{Hii} we find in W49, and thus the inferred lifetime, are smaller than previous estimations in the same region (6/12, \citealt{2001ApJ...550L..81W}) and in Sgr B2 (5/8, \citealt{2017A&A...604A..60B}). It is similar to that found in another high-mass star-forming region (1/9, \citealt{2005ApJ...624..827F}).
The period during which HMCs and H/UC\,\textsc{Hii} regions coexist in W49A (5 cores out of 17) appears to be very short, $\simeq 2\times10^4$\,yr \citep[see also the survey of][]{2021MNRAS.505.2801L}. Therefore, our results point toward a rapid dispersal of the warm, inner molecular envelope once the \textsc{Hii} region appears.

We also use the relative number of high-mass protostellar cores to H/UC\,\textsc{Hii} regions (40/85, see \cref{su:sf}) to estimate the lifetime of the massive protostellar phase, $\tau_\mathrm{proto} \simeq 40/85\,\times\,3\,10^5 \simeq 1.4\times10^5$\,yr. 
Our estimation is about two times lower than the value reported in the review of \cite{2018Motte}, $\tau_\mathrm{proto} \sim 3\times10^5$\,yr. It is also within the range of durations reported by \cite{2011ApJ...730L..33M}, $7\times10^4$\,yr to $4\times10^5$\,yr for massive YSOs with luminosities from $10^4$ to $10^5\Lsol$. 
Our estimation of the lifetime of the massive protostellar phase may be a  lower limit if we are missing detections of HMCs or H/UC\,\textsc{Hii} regions.  For example, it has been reported that a large dust optical depth could hinder the detection of molecular lines in emission
\citep[e.g.][]{2020ApJ...896L...3D}. Conversely, our estimation of the number of high-mass protostars may be an upper limit if some of the HMCs are associated with intermediate mass protostars which will not evolve to high-mass stars, or if our 3000 au-cores subfragment into several intermediate and low-mass stars.  

\section{Conclusions}
\label{s:conclu}
We presented an analysis of the star-formation activity in W49 based on the first millimeter continuum survey in this region. Our main results and conclusions are as follows:

\begin{enumerate}
    \item We built a 1.3~mm continuum image at $0.29\arcsec \times 0.24\arcsec$ resolution from ALMA B6 observations and constructed a catalog of 129 cores, extracted using the \textit{getsf} algorithm.

    \item We looked for cores associated with H/UC\,\textsc{Hii} regions by analyzing the VLA X band map (3.3~cm, \citealt{2020AJ....160..234D}), dominated by free-free emission, and the H30$\alpha$ recombination line at 231.9~GHz, covered by our ALMA observations. We found a good agreement between these two tracers of ionized gas, with 23 common detections and only four cores with \textsc{Hii} regions detected only in the X band map, out of 109 cores within the VLA field of view. 

    \item We identified a total of 36 cores associated with H/UC\,\textsc{Hii} regions over the entire ALMA mosaic, as well as four other cores tentatively associated with H/UC\,\textsc{Hii} regions. We measured their integrated free-free flux and derived their dust flux at 1.3 mm. 

    \item The spectral indexes from 3.3~cm to 1.3~mm range from 1, for the youngest cores with partially optically thick free-free emission, to about -0.1, that is optically thin free-free emission obtained for cores allegedly more evolved. The emission measures, which range from 4.5$\times10^6$ to 3.1$\times10^8$ pc\,cm$^{-6}$, are typical of H/UC\,\textsc{Hii} regions. 

    \item We also found, using the methyl formate doublet at 218.281 and 218.298~GHz within the spectral coverage of our ALMA B6 data, that 19 cores are associated with hot molecular core (HMC) emission. 17 HMCs are located within the central subregion, W49N, which is about three time the number of HMCs previously reported by \cite{2001ApJ...550L..81W}. This places W49 among the protoclusters with the best statistics for HMCs studies, along with W43 and W51\citep{2022Brouillet,2024Bonfand}.

    \item We combined these two tracers of high-mass star formation (HMC and \textsc{Hii} region emission) to propose a classification of the cores according to their evolutionary stage. Within W49N, covered by the VLA map, 69 cores without HMC or H/UC\,\textsc{Hii} region are labeled as "young cores", meaning that they are either prestellar or lower-mass protostellar cores.
    The 5 cores with H/UC\,\textsc{Hii} regions which are also associated with HMCs are assumed to be less evolved than the 23 cores with H/UC\,\textsc{Hii} regions only. Finally we also considered a fifth category of H/UC\,\textsc{Hii} regions not associated to dust cores. 

    \item The statistical lifetimes of the hot molecular core and massive protostellar phases in W49 are estimated to be $6\times10^4$\,yr and $1.4\times10^5$\,yr, respectively, based on a duration of the H/UC\,\textsc{Hii} phase of $3\times10^5$\,yr. The identification of massive protostellar cores is based on the association between a dust core and a HMC and/or an H/UC\,\textsc{Hii} region. These estimations could be upper limits taking into the possible subfragmentation of HMCs into less massive protostars. 

    \item We found that HMCs and H/UC\,\textsc{Hii} regions coexist in W49A during a short period of $\simeq 2\times10^4$\,yr. This indicates a rapid dispersal of the inner molecule envelope once the HC\,\textsc{Hii} is formed. 
    
\end{enumerate}

\begin{acknowledgements}
We thank the referee R. Miyawaki for his helpful comments improving the manuscript. 
RGM and TN acknowledge support from UNAM-PAPIIT project IN108822 and from CONACyT Ciencia de Frontera project ID 86372. TN also acknowledges support from the postdoctoral fellowship program of the UNAM. Part of this work was performed using the high-performance computers at IRyA, Mexico, funded by CONACyT and UNAM. The work from the IT staff at this institute is acknowledged. 
AG acknowledges support from the NSF via grants AST 2008101 and CAREER 2142300.
H.B.L. is supported by the National Science and Technology Council (NSTC) of Taiwan (Grant Nos. 111-2112-M-110-022-MY3).
C R-Z acknowledges support from program UNAM-PAPIIT IG101723.
This paper makes use of the following ALMA data: ADS/JAO.ALMA\#2016.1.00620.S ALMA is a partnership of ESO (representing its member states), NSF (USA) and NINS (Japan), together with NRC (Canada), MOST and ASIAA (Taiwan), and KASI (Republic of Korea), in cooperation with the Republic of Chile. The Joint ALMA Observatory is operated by ESO, AUI/NRAO and NAOJ. This work is based on an analysis carried out with the GILDAS, IMAGER and CASSIS softwares, as well as the CDMS and JPL spectroscopic databases. CASSIS has been developed by IRAP-UPS/CNRS (http://cassis.irap.omp.eu).
\end{acknowledgements}

\bibliographystyle{aa} 
\bibliography{bib_W49}

\begin{appendix}
\renewcommand{\thefigure}{A\arabic{figure}}
\renewcommand{\thetable}{A\arabic{table}}

\section{Core catalog}
\cref{tab:cores-par} lists the physical properties, detection (or non detection) of H/UC\,\textsc{Hii} regions and hot molecular cores for the 129 cores detected at 1.3~mm in W49. The labels of H/UC\,\textsc{Hii} regions previously identified at the core positions are also given.

\onecolumn 
\setlength{\LTcapwidth}{5in}

\begin{ThreePartTable}
  \begin{TableNotes}  
  \small
  \item (1) Core number in the getsf catalog. (2) Right Ascension and Declination coordinates. (3) Subregion: (W49-)N, (W49-)SW or (W49)-out(skirts). (4) FWHM major and minor axis. (5) Position Angle (West to North) of the ellipse. (6,7) Integrated fluxes in B6 and BX. (8) Detection of the H/UC\,\textsc{Hii} region in X band 3.3~cm (BX) and/or in H30$\alpha$ (RL). (9) Identification of the associated H/UC\,\textsc{Hii} region in the catalogs of \cite{1997ApJ...482..307D,2020AJ....160..234D}. (10) Cores associated to hot molecular core emission are marked with *. 
  \end{TableNotes}

\begin{longtable}{ccccc|cccccc} 
\caption{Main characteristics of cores detected on the 1.3~mm continuum image.} 
\label{tab:cores-par} 
\\ 
\hline 
\hline
n & R.A. Dec. & Region & Size & $PA$ & $F^{\rm int}_{\rm B6}$ & $F^{\rm int}_{\rm BX}$ & \textsc{Hii} & Cross ID & HMC  \\ 
 & [ICRS] & & [$\arcsec \times \arcsec$] & & [mJy.beam$^{-1}$] & [mJy.beam$^{-1}$] & & & \\ 
 (1) & (2) & (3) & (4) & (5) & (6) & (7) & (8) & (9) & (10) \\
\hline 
\endfirsthead

\multicolumn{10}{c}{{\textbf{\tablename\ \thetable{}} -- continued from previous page}}\\
\hline
\hline
n & R.A. Dec. & Region & Size & $PA$ & $F^{\rm int}_{\rm B6}$ & $F^{\rm int}_{\rm BX}$ & \textsc{Hii} &  Cross ID & HMC  \\ 
 & [ICRS] & & [$\arcsec \times \arcsec$] & & [mJy.beam$^{-1}$] & [mJy.beam$^{-1}$] & & & \\ 
 (1) & (2) & (3) & (4) & (5) & (6) & (7) & (8) & (9) & (10) \\
\hline 
\endhead

\hline
\insertTableNotes         
\endfoot

\endlastfoot
  1 & 19:10:13.419~~9:06:13.00 &   N & 0.33$\,\times\,$0.27 &  10 & 1031.9 &  36.6 & BX+RL & G2a &   & \\ 
  2 & 19:10:13.149~~9:06:12.72 &   N & 0.37$\,\times\,$0.33 & 117 &  728.4 &  38.5 & BX+RL & B2  &   & \\ 
  3 & 19:10:12.889~~9:06:11.73 &   N & 0.39$\,\times\,$0.29 & 169 &  365.0 &  56.0 & BX+RL & A2  &   & \\ 
  4 & 19:10:10.938~~9:05:17.53 &  SW & 0.31$\,\times\,$0.27 & 166 &   94.6 &       &    RL &  -  & * & \\ 
  5 & 19:10:16.356~~9:06:07.04 & out & 0.49$\,\times\,$0.35 & 115 &  133.2 &       &    RL &  O  &   & \\ 
  6 & 19:10:13.614~~9:06:49.12 &   N & 0.45$\,\times\,$0.30 & 118 &   75.0 &   -   &       &     & * & \\ 
  7 & 19:10:13.144~~9:06:18.76 &   N & 0.35$\,\times\,$0.29 & 142 &  188.5 &  40.6 & BX+RL &  C  & * & \\ 
  8 & 19:10:14.130~~9:06:25.00 &   N & 0.44$\,\times\,$0.30 & 155 &  200.2 &   1.5 &    BX &  J1 & * & \\ 
  9 & 19:10:11.050~~9:05:20.11 &  SW & 0.44$\,\times\,$0.40 &  17 &   85.9 &       &    RL &  R  & * & \\ 
 10 & 19:10:13.455~~9:06:12.79 &   N & 0.33$\,\times\,$0.27 &  20 &  332.0 &  12.3 & BX+RL & G2c &   & \\ 
 11 & 19:10:15.369~~9:06:14.95 &   N & 0.40$\,\times\,$0.33 & 174 &   56.2 &  22.0 & BX+RL &  N  &   & \\ 
 12 & 19:10:16.372~~9:06:06.81 & out & 0.47$\,\times\,$0.31 & 123 &  116.1 &       &    RL &  O  &   & \\ 
 13 & 19:10:13.116~~9:06:12.36 &   N & 0.30$\,\times\,$0.26 &  11 &  149.0 &  12.7 & BX+RL &  B1 &   & \\ 
 14 & 19:10:13.202~~9:06:11.13 &   N & 0.68$\,\times\,$0.57 &  69 &  203.9 & 135.2 & BX+RL &  D  &   & \\ 
 15 & 19:10:13.340~~9:06:21.37 &   N & 0.56$\,\times\,$0.47 &  97 &   59.1 &  60.1 & BX+RL &  F  &   & \\ 
 16 & 19:10:13.373~~9:06:12.58 &   N & 0.46$\,\times\,$0.39 & 146 &  279.9 &  69.0 & BX+RL &  G1 &   & \\ 
 17 & 19:10:12.884~~9:06:12.16 &   N & 0.39$\,\times\,$0.34 &  30 &  125.9 &  17.7 & BX+RL &  A1 & * & \\ 
 18 & 19:10:13.312~~9:06:19.87 &   N & 0.44$\,\times\,$0.32 & 172 &   93.3 &   1.0 &    BX &  -  &   & \\ 
 19 & 19:10:14.645~~9:06:25.97 &   N & 0.41$\,\times\,$0.40 & 143 &   20.3 &   -   &       &     & * & \\ 
 20 & 19:10:13.279~~9:06:12.79 &   N & 0.39$\,\times\,$0.29 &   0 &   82.6 &   -   &       &     & * & \\ 
 22 & 19:10:14.628~~9:06:45.96 &   N & 0.38$\,\times\,$0.33 &   6 &   10.5 &   -   &       &     &   & \\ 
 23 & 19:10:13.060~~9:06:16.07 &   N & 0.35$\,\times\,$0.33 & 178 &   17.7 &  12.7 & BX+RL &  C1 &   & \\ 
 24 & 19:10:14.284~~9:06:22.22 &   N & 0.44$\,\times\,$0.40 &  69 &   20.7 &   -   &       &     &   & \\ 
 25 & 19:10:15.358~~9:06:15.36 &   N & 0.36$\,\times\,$0.33 & 167 &   21.3 &   -   &       &     & * & \\ 
 26 & 19:10:12.665~~9:06:07.57 &   N & 0.48$\,\times\,$0.44 &  71 &   25.0 &   -   &       &     &   & \\ 
 27 & 19:10:13.272~~9:06:16.38 &   N & 0.52$\,\times\,$0.37 &  47 &   53.7 &   -   &       &     & * & \\ 
 28 & 19:10:13.149~~9:06:13.84 &   N & 0.45$\,\times\,$0.38 &  72 &  112.0 &   -   &       &     & * & \\ 
 29 & 19:10:13.259~~9:06:12.19 &   N & 0.37$\,\times\,$0.31 &  35 &   23.3 &  17.3 & BX+RL &  E3 &   & \\ 
 30 & 19:10:14.137~~9:06:26.60 &   N & 0.45$\,\times\,$0.38 & 110 &   27.9 &   -   &       &     &   & \\ 
 31 & 19:10:14.312~~9:06:22.21 &   N & 0.37$\,\times\,$0.27 &  23 &   11.7 &   -   &       &     &   & \\ 
 32 & 19:10:13.170~~9:06:13.36 &   N & 0.36$\,\times\,$0.33 &  64 &   70.4 &   -   &       &     & * & \\ 
 34 & 19:10:13.259~~9:06:18.88 &   N & 0.38$\,\times\,$0.29 & 164 &   32.2 &   -   &       &     &   & \\ 
 35 & 19:10:13.441~~9:06:07.11 &   N & 0.28$\,\times\,$0.25 &  27 &    4.5 &   -   &       &     &   & \\ 
 36 & 19:10:12.097~~9:06:14.64 &   N & 0.41$\,\times\,$0.26 &  74 &    9.0 &   -   &       &     & * & \\ 
 37 & 19:10:14.182~~9:06:20.58 &   N & 0.32$\,\times\,$0.28 & 103 &    2.9 &   -   &       &     &   & \\ 
 39 & 19:10:11.671~~9:05:26.86 &  SW & 0.55$\,\times\,$0.45 & 153 &   30.6 &       &    RL &  S  &   & \\ 
 40 & 19:10:13.392~~9:06:11.82 &   N & 0.56$\,\times\,$0.48 &  10 &   79.0 &  54.8 & BX+RL & G1S &   & \\ 
 42 & 19:10:14.664~~9:06:25.53 &   N & 0.45$\,\times\,$0.36 & 107 &   12.0 &   -   &       &     &   & \\ 
 43 & 19:10:13.348~~9:06:16.05 &   N & 0.39$\,\times\,$0.33 & 155 &   25.9 &   -   &       &     & * & \\ 
 44 & 19:10:12.703~~9:06:11.29 &   N & 0.31$\,\times\,$0.26 &  24 &   22.2 &   1.5 & BX+RL &  -  & * & \\ 
 45 & 19:10:13.211~~9:06:14.30 &   N & 0.31$\,\times\,$0.27 &  26 &   24.8 &   -   &       &     &   & \\ 
 46 & 19:10:13.498~~9:06:11.77 &   N & 0.36$\,\times\,$0.32 & 179 &   58.1 &  34.7 & BX+RL & G3b &   & \\ 
 47 & 19:10:13.498~~9:06:12.42 &   N & 0.37$\,\times\,$0.26 &   8 &   43.7 &  20.3 & BX+RL & G3a &   & \\ 
 48 & 19:10:13.721~~9:06:44.58 &   N & 0.37$\,\times\,$0.33 &  53 &    3.6 &   -   &       &     &   & \\ 
 49 & 19:10:12.671~~9:06:18.28 &   N & 0.39$\,\times\,$0.29 & 172 &    7.0 &   -   &       &     &   & \\ 
 50 & 19:10:10.686~~9:05:16.78 &  SW & 0.95$\,\times\,$0.70 & 134 &   36.7 &       &    RL &  R3 &   & \\ 
 51 & 19:10:12.644~~9:06:18.93 &   N & 0.39$\,\times\,$0.30 &  37 &    4.0 &   -   &       &     &   & \\ 
 52 & 19:10:16.504~~9:06:04.24 & out & 0.63$\,\times\,$0.53 &  23 &   11.4 &       &       &     &   & \\ 
 53 & 19:10:13.364~~9:06:18.01 &   N & 0.44$\,\times\,$0.41 &  88 &   10.3 &   -   &       &     &   & \\ 
 54 & 19:10:13.323~~9:06:16.32 &   N & 0.33$\,\times\,$0.30 &  27 &   16.8 &   -   &       &     & * & \\ 
 55 & 19:10:13.230~~9:06:54.16 &   N & 0.72$\,\times\,$0.58 & 130 &    4.1 &   -   &       &     &   & \\ 
 56 & 19:10:13.188~~9:06:12.85 &   N & 0.45$\,\times\,$0.36 &   0 &   81.5 &   -   &       &     & * & \\ 
 57 & 19:10:13.114~~9:06:13.77 &   N & 0.42$\,\times\,$0.37 & 120 &   62.3 &   -   &       &     &   & \\ 
 58 & 19:10:16.666~~9:05:49.88 & out & 0.44$\,\times\,$0.32 &  53 &    7.2 &       &       &     &   & \\ 
 59 & 19:10:16.402~~9:06:07.36 & out & 0.47$\,\times\,$0.36 & 143 &   22.8 &       &    RL &  -  &   & \\ 
 60 & 19:10:15.754~~9:06:05.78 &   N & 0.49$\,\times\,$0.40 & 169 &    4.7 &   -   &       &     &   & \\ 
 61 & 19:10:15.256~~9:06:10.99 &   N & 0.32$\,\times\,$0.29 &  92 &    2.1 &   -   &       &     &   & \\ 
 62 & 19:10:12.843~~9:06:11.44 &   N & 0.61$\,\times\,$0.42 & 149 &   76.7 &   -   &       &     &   & \\ 
 63 & 19:10:14.605~~9:06:26.38 &   N & 0.45$\,\times\,$0.32 &   4 &    7.2 &   -   &       &     &   & \\ 
 65 & 19:10:14.159~~9:06:21.56 &   N & 0.60$\,\times\,$0.54 & 110 &   14.2 &   -   &       &     &   & \\ 
 66 & 19:10:13.385~~9:06:19.37 &   N & 0.40$\,\times\,$0.36 &  37 &   19.7 &   -   &       &     &   & \\ 
 67 & 19:10:13.656~~9:06:21.49 &   N & 0.57$\,\times\,$0.43 &  46 &   10.0 &   -   &       &     &   & \\ 
 68 & 19:10:12.800~~9:06:27.54 &   N & 0.54$\,\times\,$0.34 &  85 &    6.9 &   -   &       &     &   & \\ 
 69 & 19:10:14.186~~9:06:15.32 &   N & 0.66$\,\times\,$0.41 &  42 &   21.3 &  17.2 & BX+RL &  J  &   & \\ 
 70 & 19:10:12.746~~9:06:10.33 &   N & 0.36$\,\times\,$0.32 &  53 &   24.0 &   -   &       &     &   & \\ 
 71 & 19:10:13.434~~9:06:22.78 &   N & 0.46$\,\times\,$0.36 & 155 &    7.1 &   -   &       &     &   & \\ 
 72 & 19:10:13.178~~9:06:18.28 &   N & 0.50$\,\times\,$0.40 & 143 &   42.5 &  31.2 & BX+RL &  C  &   & \\ 
 73 & 19:10:13.601~~9:06:11.02 &   N & 0.43$\,\times\,$0.30 &   0 &   45.6 &  30.1 & BX+RL & G3d &   & \\ 
 74 & 19:10:14.563~~9:06:20.47 &   N & 0.50$\,\times\,$0.40 &  50 &    8.7 &   3.6 & BX+RL &  L  &   & \\ 
 75 & 19:10:14.126~~9:06:24.28 &   N & 0.33$\,\times\,$0.21 &  20 &   17.8 &   -   &       &     &   & \\ 
 76 & 19:10:13.323~~9:06:25.53 &   N & 0.47$\,\times\,$0.31 & 124 &    6.0 &   -   &       &     &   & \\ 
 77 & 19:10:13.898~~9:06:36.91 &   N & 0.29$\,\times\,$0.24 &  47 &    1.5 &   -   &       &     &   & \\ 
 78 & 19:10:16.344~~9:06:07.75 & out & 0.53$\,\times\,$0.49 & 111 &   21.8 &       &    RL &  -  &   & \\ 
 79 & 19:10:13.018~~9:06:17.05 &   N & 0.60$\,\times\,$0.55 &  90 &   16.2 &   -   &       &     &   & \\ 
 80 & 19:10:10.801~~9:05:19.45 &  SW & 1.03$\,\times\,$0.77 & 137 &   25.9 &       &       &     &   & \\ 
 81 & 19:10:14.610~~9:06:02.65 &   N & 0.47$\,\times\,$0.37 &  41 &    6.0 &   -   &       &     &   & \\ 
 82 & 19:10:14.178~~9:06:26.65 &   N & 0.34$\,\times\,$0.29 &  55 &    9.1 &   -   &       &     &   & \\ 
 83 & 19:10:13.135~~9:06:13.45 &   N & 0.43$\,\times\,$0.38 &  75 &   83.7 &   -   &       &     &   & \\ 
 84 & 19:10:12.329~~9:06:18.58 &   N & 0.28$\,\times\,$0.25 & 176 &    2.0 &   -   &       &     &   & \\ 
 85 & 19:10:12.720~~9:06:28.69 &   N & 0.30$\,\times\,$0.28 &  18 &    2.8 &   -   &       &     &   & \\ 
 86 & 19:10:13.234~~9:06:12.68 &   N & 0.44$\,\times\,$0.40 & 177 &   49.8 &   -   &       &     &   & \\ 
 87 & 19:10:14.184~~9:06:14.37 &   N & 1.05$\,\times\,$0.94 & 178 &   32.2 &  13.9 & BX+RL &  J  &   & \\ 
 88 & 19:10:12.613~~9:06:22.60 &   N & 0.43$\,\times\,$0.31 & 168 &    3.5 &   -   &       &     &   & \\ 
 89 & 19:10:13.231~~9:06:08.87 &   N & 0.53$\,\times\,$0.38 & 152 &    9.6 &   -   &       &     &   & \\ 
 90 & 19:10:15.433~~9:06:16.78 &   N & 0.43$\,\times\,$0.32 &  11 &    3.3 &   -   &       &     &   & \\ 
 91 & 19:10:11.174~~9:05:53.85 &   N & 0.45$\,\times\,$0.34 &  12 &    3.1 &   -   &       &     &   & \\ 
 92 & 19:10:12.685~~9:06:32.36 &   N & 0.97$\,\times\,$0.90 & 127 &    7.3 &   -   &       &     &   & \\ 
 93 & 19:10:15.317~~9:06:14.80 &   N & 0.54$\,\times\,$0.34 &   5 &   13.4 &   -   &       &     &   & \\ 
 94 & 19:10:14.509~~9:06:22.47 &   N & 1.06$\,\times\,$0.82 &  67 &   25.7 &   -   &    RL &  L  &   & \\ 
 95 & 19:10:11.487~~9:05:32.14 &   N & 0.31$\,\times\,$0.27 &  24 &    2.4 &   -   &       &     &   & \\ 
 96 & 19:10:13.571~~9:06:21.34 &   N & 0.47$\,\times\,$0.35 & 173 &    9.9 &   -   &       &     &   & \\ 
 97 & 19:10:13.307~~9:06:17.02 &   N & 0.37$\,\times\,$0.29 & 162 &   15.2 &   -   &       &     &   & \\ 
 98 & 19:10:14.796~~9:06:44.25 &   N & 0.68$\,\times\,$0.47 &  81 &    5.8 &   -   &       &     &   & \\ 
 99 & 19:10:15.807~~9:06:05.59 &   N & 0.29$\,\times\,$0.27 & 157 &    1.4 &   -   &       &     &   & \\ 
100 & 19:10:11.490~~9:06:24.34 &   N & 0.35$\,\times\,$0.29 &  11 &    1.5 &   -   &       &     &   & \\ 
101 & 19:10:13.459~~9:06:20.14 &   N & 0.49$\,\times\,$0.32 & 144 &   15.4 &   -   &       &     &   & \\ 
102 & 19:10:12.779~~9:06:11.13 &   N & 0.34$\,\times\,$0.29 &  24 &   19.5 &   -   &       &     &   & \\ 
103 & 19:10:14.825~~9:06:36.88 &   N & 0.69$\,\times\,$0.50 & 140 &    5.1 &   -   &       &     &   & \\ 
104 & 19:10:14.037~~9:06:23.56 &   N & 0.41$\,\times\,$0.35 & 122 &   15.5 &   0.7 &    BX &  -  & * & \\ 
105 & 19:10:13.630~~9:06:20.30 &   N & 1.03$\,\times\,$0.86 & 146 &   33.5 &   -   &       &     &   & \\ 
106 & 19:10:12.720~~9:06:28.38 &   N & 0.41$\,\times\,$0.39 &  52 &    4.5 &   -   &       &     &   & \\ 
107 & 19:10:16.297~~9:06:06.66 & out & 0.55$\,\times\,$0.40 & 157 &   14.8 &       &    RL &  -  &   & \\ 
108 & 19:10:13.719~~9:06:25.03 &   N & 0.40$\,\times\,$0.34 & 111 &    5.1 &   -   &       &     &   & \\ 
109 & 19:10:14.237~~9:06:28.51 &   N & 0.48$\,\times\,$0.37 & 162 &   11.0 &   -   &       &     &   & \\ 
110 & 19:10:15.409~~9:06:11.87 &   N & 0.65$\,\times\,$0.36 &   2 &    5.8 &   -   &       &     &   & \\ 
111 & 19:10:11.629~~9:06:16.03 &   N & 0.35$\,\times\,$0.26 & 172 &    3.4 &   -   &       &     &   & \\ 
112 & 19:10:11.731~~9:05:27.37 &  SW & 0.67$\,\times\,$0.43 &  10 &   25.7 &       &    RL &  S  &   & \\ 
113 & 19:10:10.277~~9:05:14.07 &  SW & 0.30$\,\times\,$0.23 & 160 &    4.5 &       &       &     &   & \\ 
114 & 19:10:12.838~~9:06:29.46 &   N & 0.34$\,\times\,$0.27 & 180 &    3.0 &   -   &       &     &   & \\ 
115 & 19:10:16.826~~9:05:49.99 & out & 0.83$\,\times\,$0.66 &   4 &   11.1 &       &       &     &   & \\ 
116 & 19:10:14.747~~9:06:25.65 &   N & 0.50$\,\times\,$0.34 & 150 &    4.9 &   3.4 & BX+RL &  M  &   & \\ 
117 & 19:10:13.683~~9:06:10.06 &   N & 0.36$\,\times\,$0.34 & 119 &    6.4 &   2.4 &    BX &  -  &   & \\ 
119 & 19:10:10.956~~9:05:18.60 &  SW & 0.34$\,\times\,$0.25 &  90 &    2.5 &       &       &     &   & \\ 
120 & 19:10:14.284~~9:06:32.28 &   N & 0.55$\,\times\,$0.53 &  52 &   11.3 &   -   &       &     &   & \\ 
121 & 19:10:14.253~~9:06:29.98 &   N & 0.42$\,\times\,$0.35 & 115 &    4.0 &   -   &       &     &   & \\ 
122 & 19:10:12.416~~9:06:06.76 &   N & 0.75$\,\times\,$0.64 &  72 &    8.4 &   -   &       &     &   & \\ 
123 & 19:10:10.758~~9:05:21.46 &  SW & 0.30$\,\times\,$0.23 &  34 &    1.7 &       &       &     &   & \\ 
124 & 19:10:13.735~~9:06:49.42 &   N & 0.44$\,\times\,$0.40 & 147 &    4.4 &   -   &       &     &   & \\ 
125 & 19:10:14.685~~9:06:45.97 &   N & 0.33$\,\times\,$0.24 &  12 &    2.4 &   -   &       &     &   & \\ 
126 & 19:10:13.779~~9:06:16.81 &   N & 0.38$\,\times\,$0.26 &  15 &    3.5 &   -   &       &     &   & \\ 
128 & 19:10:16.708~~9:05:49.17 & out & 0.34$\,\times\,$0.27 &  73 &    3.4 &       &       &     &   & \\ 
130 & 19:10:13.780~~9:06:22.11 &   N & 0.37$\,\times\,$0.29 & 173 &    4.1 &   -   &       &     &   & \\ 
131 & 19:10:10.541~~9:05:14.08 &  SW & 0.30$\,\times\,$0.25 & 124 &    3.2 &       &    RL &  Q  &   & \\ 
133 & 19:10:14.122~~9:06:26.01 &   N & 0.39$\,\times\,$0.35 &  65 &    9.9 &   -   &       &     & * & \\ 
138 & 19:10:14.525~~9:06:28.78 &   N & 0.40$\,\times\,$0.30 & 173 &    1.7 &   -   &       &     &   & \\ 
140 & 19:10:14.262~~9:06:26.22 &   N & 0.36$\,\times\,$0.26 & 117 &    3.0 &   -   &       &     &   & \\ 
141 & 19:10:13.980~~9:06:24.90 &   N & 0.34$\,\times\,$0.27 & 103 &    2.9 &   -   &       &     &   & \\ 
142 & 19:10:14.980~~9:06:03.82 &   N & 0.30$\,\times\,$0.23 & 142 &    1.7 &   -   &       &     &   & \\ 
147 & 19:10:10.893~~9:05:16.78 &  SW & 0.95$\,\times\,$0.68 & 152 &   20.0 &       &    RL &  -  &   & \\ 
\hline    
   
\end{longtable}    
\end{ThreePartTable}   
   
\twocolumn

\end{appendix}

\end{document}